\documentclass[preprint]{aastex}

\newcommand{\nh}{\mbox{$N_{\rm H}$}}

\newcommand{\HII}{\ion{H}{2}}
\newcommand{\skipthis}[1]{}

\newcommand{\psqcm}{{\rm cm}^{-2}}
\newcommand{\persqcm}{\rm \,cm^{-2}}

\newcommand{\ps}{{\rm s}^{-1}}

\newcommand{\erg}{{\rm ergs}}

\def\micron{\hbox{$\mu$m}}
\newcommand{\be}{\begin{equation}}
\newcommand{\ee}{\end{equation}}
\newcommand{\e}{et al.\ }

\slugcomment{\today}

\shorttitle{RCW 38 Point Sources}
\shortauthors{Wolk et al.}

\begin{document}


\title{X-ray and IR Point Source Identification and Characteristics \\
    In the Embedded, Massive Star-Forming Region RCW 38}


\author{Scott J. Wolk,
Bradley D. Spitzbart,
Tyler L. Bourke}
\affil{Harvard--Smithsonian Center for Astrophysics,
       60 Garden Street, Cambridge, MA 02138}

\and
\author{Jo\~ao Alves \altaffilmark{1}}
\affil{European Southern Observatory, Karl-Schwarzschild Strasse 2,
        D-85748 Garching bei M\"unchen, Germany}


\altaffiltext{1}{Current Address: Calar Alto Observatory, Centro Astron\'omico
Hispano Alem\'an C/Jes\'us Durbán Rem\'on 2-2 E-04004 Almería, Spain }


\begin{abstract}
We report on results of a 96.7 ks {\it Chandra} observation of 
one of the youngest, most embedded and massive young stellar 
clusters studied in X-rays -- RCW~38.  We detect 460 
sources in the field of which 360 are confirmed to be 
associated with the RCW~38
cluster. The cluster members range in 
luminosity from $~10^{30}$ ergs s$^{-1}$
to $ 10^{33.5}$ ergs s$^{-1}$. 
Over ten percent of the cluster members
with over 100 counts  exhibit flares
while about 15\% of cluster members with over 30 counts are variable.  
Of the sources identified as cluster members, 160 have near-infrared
(NIR) counterparts either in the 2MASS database or detected via VLT 
observations.  Of these about 20\% appear to have optically thick 
disks.  An additional 353 members are identified through NIR
observations of which at least
50\% possess optically thick disks.
We fit over 100 X-ray sources as absorbed Raymond-Smith type plasmas and 
find the column to the cluster members varies from $10^{21.5}$ to 
$10^{23}\persqcm$.  We compare the gas to dust absorption signatures
in these stars and find \nh= A$_V \times 2 \times 10^{21}$cm$^{-2}$.
We find that the cluster contains 
31 candidate OB stars and is centered about 
10\arcsec\ (0.1 pc) west of the primary source of the ionization, the
O5 star IRS~2.  The cluster has a peak central density 
of about 400 X-ray sources pc $^{-2}$.   We
estimate that the total cluster membership exceeds 2000 stars. 
\end{abstract}




\section{Introduction}
The evolution of high mass clustered star forming regions is complex and poorly understood.  Only the nearby (0.5 kpc), optically revealed, Orion Nebular 
Cluster (ONC) is well studied.  Yet, a wide variety of high mass embedded clusters are found within 2 kpc of the Sun (Lada \& Lada 2003).  Within this limit, the young cluster RCW~38 (l=268$^{\rm o}$, b=-1$^{\rm o}$) is unique as the only region other than the ONC to contain over 1300 members. RCW~38 is significantly more embedded and spatially denser that the ONC (Smith \e 1999) and other regions that have been studied recently with $Spitzer$ and $Chandra$ 
(e.g., Trifid (Rho \e 2006), M17 and the Rosette Nebula (Townsley \e 2003),  RCW~49 (Whitney \e 2004, Churchwell \e 2004)). RCW~38 provides a unique opportunity to study the evolution of a rich cluster during the phase where its most massive member (O5) has just completed its ultracompact HII region (UCHII) phase and is now greatly influencing its natal environment and the evolution of its low mass members.

At a distance of 1.7 kpc, \objectname{RCW~38} (Rodgers, Campbell \& Whiteoak 
1960) is one of the brightest \ion{H}{2}\ regions at radio wavelengths (e.g., Wilson \e 1970).  It is a uniquely young ($<$1 Myr), embedded ($A_V \sim 10$) stellar cluster surrounding an early O star, IRS~2 ($\sim$O5; Frogel \& Persson 
1974; Smith \e 1999), $Chandra$ and deep near infrared observations reveal a dense cluster embedded in a diffuse hot plasma (Wolk \e 2002). The O star has evacuated its immediate surroundings of dust, creating a wind bubble $\sim$0.1 pc in radius (Smith \e 1999, Vigil \e in prep.) which is confined by the surrounding molecular cloud, as traced by mm continuum and molecular line emission (Bourke \e 2004).  This bubble is filled with diffuse thermal and synchrotron X-ray emission (Wolk \e 2002), and diffuse emission can be traced outside of the bubble, showing a similar structure to the diffuse infrared emission in the region. At the interface between the bubble and cloud is a region of warm dust and ionized gas, which shows evidence for ongoing star formation, particularly along its western edge (Smith \e 1999, Vigil \e in prep.).  Extended warm dust is found throughout the 2--3 pc region at mid-infrared wavelengths, and coincides with the extended X-ray plasma. This is evidence that the influence of the massive stars reaches beyond the confines of the O star bubble. RCW~38 appears similar in structure to RCW~49 and M~20 which have been studied with IRAC and MIPS on $Spitzer$ (Whitney \e 2004, Churchwell \e 2004 and Rho \e 2006) but it is not as evolved. RCW~38 appears to be a blister compact \ion{H}{2}\ region lying just inside the edge of a giant molecular cloud.  

In this paper we discuss the X-ray point sources observed in our recent $Chandra$ and near-infrared observations.  Section 2 describes the basic observational setup and data reduction, \S 3 assesses cluster membership via quartile (X-ray color) analysis, \S 4 examines X-ray source variability while \S 5 models the X-ray spectra.  Both of these results are used to evaluate the effectiveness of using X-ray color as a membership criterion.  In \S 6 we fold in ground based infrared data for a variety of purposes.  We estimate disk parameters for the X-ray sources as well as evaluating the full stellar content of the cluster. We further evaluate the relationship between the inferred gas and dust columns along the line of sight and examine the effect of X-ray biases in source selection.  We also estimate the total size of the cluster with special focus on the O and B stars.  In \S 7 we pay special attention to some extreme objects in the field and their implications for the 3 dimensional structure of the region.  We summarize our results in \S 8.   

\section{Observation and Data Reduction} 
{\it Chandra} observed RCW 38 on 2001 December 10-11 for 96.7 ks (ObsID
2556), using Advanced CCD Imaging Spectrometer (ACIS) chips 0, 1, 2, 3, 6 and 7 in very faint mode. The ACIS combines the ability to image X-rays at a 0\farcs492 plate scale with moderate spectral resolution (R$\sim 30-50$) and time tagging every 3.2 seconds. The combined field of view (FOV) of the 4 chips in the imaging array is 16\farcm9 $\times$ 16\farcm9. The aimpoint of the array was 8h59m19.20s $-47^{\rm o}30\arcmin22\farcs0 $ (J2000.0), and the satellite roll angle (i.e., orientation of the CCD array relative to the north--south direction) was 51$^{\rm o}$. The roll angle was selected so that photons from the \HII\ region Bran~213A would strike the ACIS spectroscopic array where upon chips 6 and 7 were also turned on.  While the \HII\ region was clearly detected these data will not be discussed here since the mirror point-spread function (PSF) is considerably degraded far off-axis.  The focal plane temperature was -119.7$^{\rm o}$C.  The radiation environment was benign.

\subsection{Data Preparation}
The raw data was originally processed under the $Chandra$ X-ray Center's standard processing version 6.4.0.  We reprocessed the observation using the latest version 6.13.2.  Using acis\_process\_events on the level 1 events files, a gain correction is applied from CALDB 2.21.  The CTI correction and VFAINT background cleaning, available since version 6.12, were also applied.
An energy filter was applied to remove photons above 8 keV and below 300 eV. We found that the VFAINT correction removed good events from bright sources and did not reveal any new faint sources.  So the VFAINT correction was backed out leaving only the CTI correction. Following standard processing, bad grade and bad status (grades 1, 5 and 7, status $>$0) events and bad time intervals were filtered using the CIAO tool {\it dmcopy}. 

\begin{figure}[h]
\plotfiddle{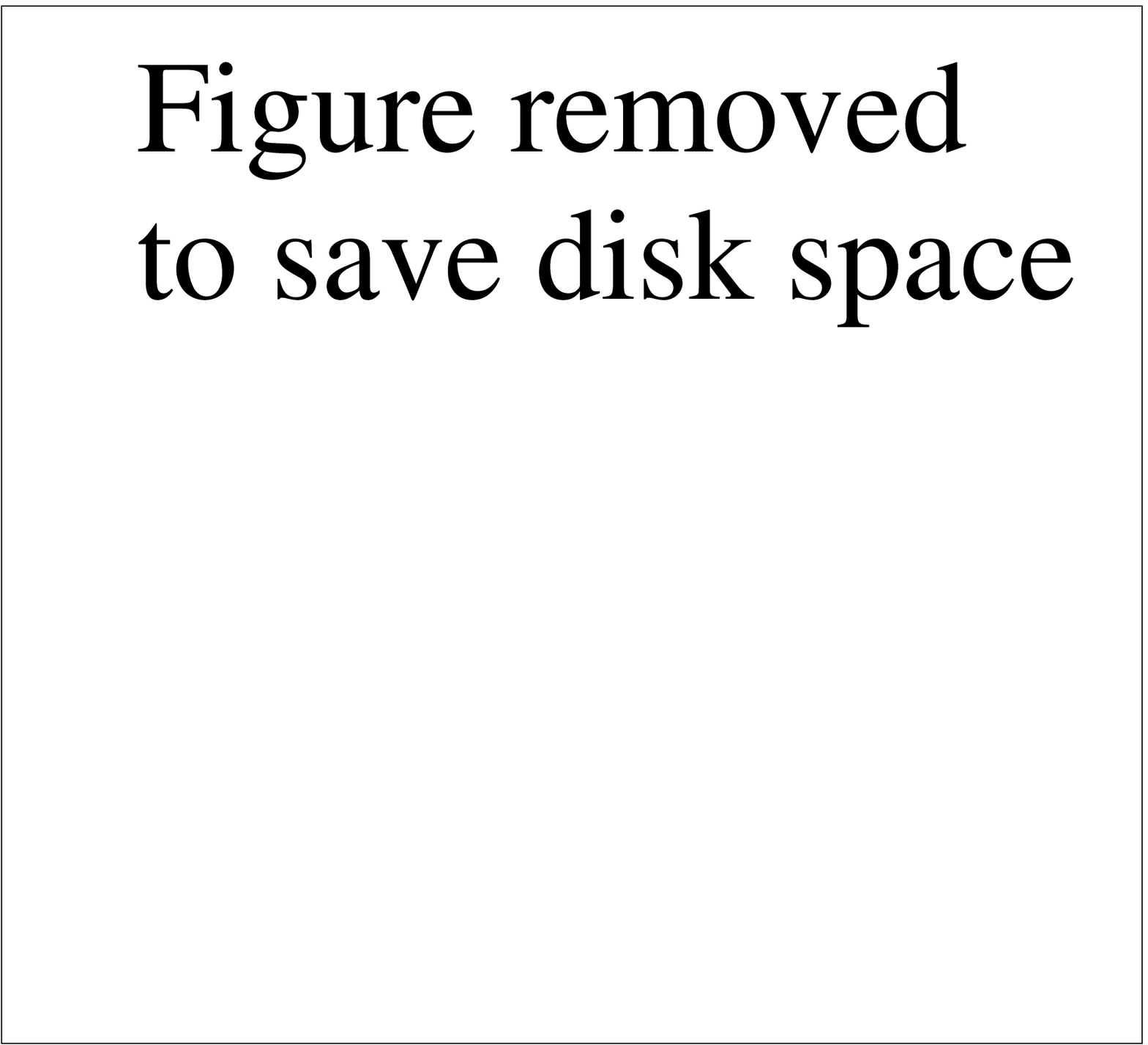}{0.0in}{0}{300.}{265.}{50.}{0.}
\plotfiddle{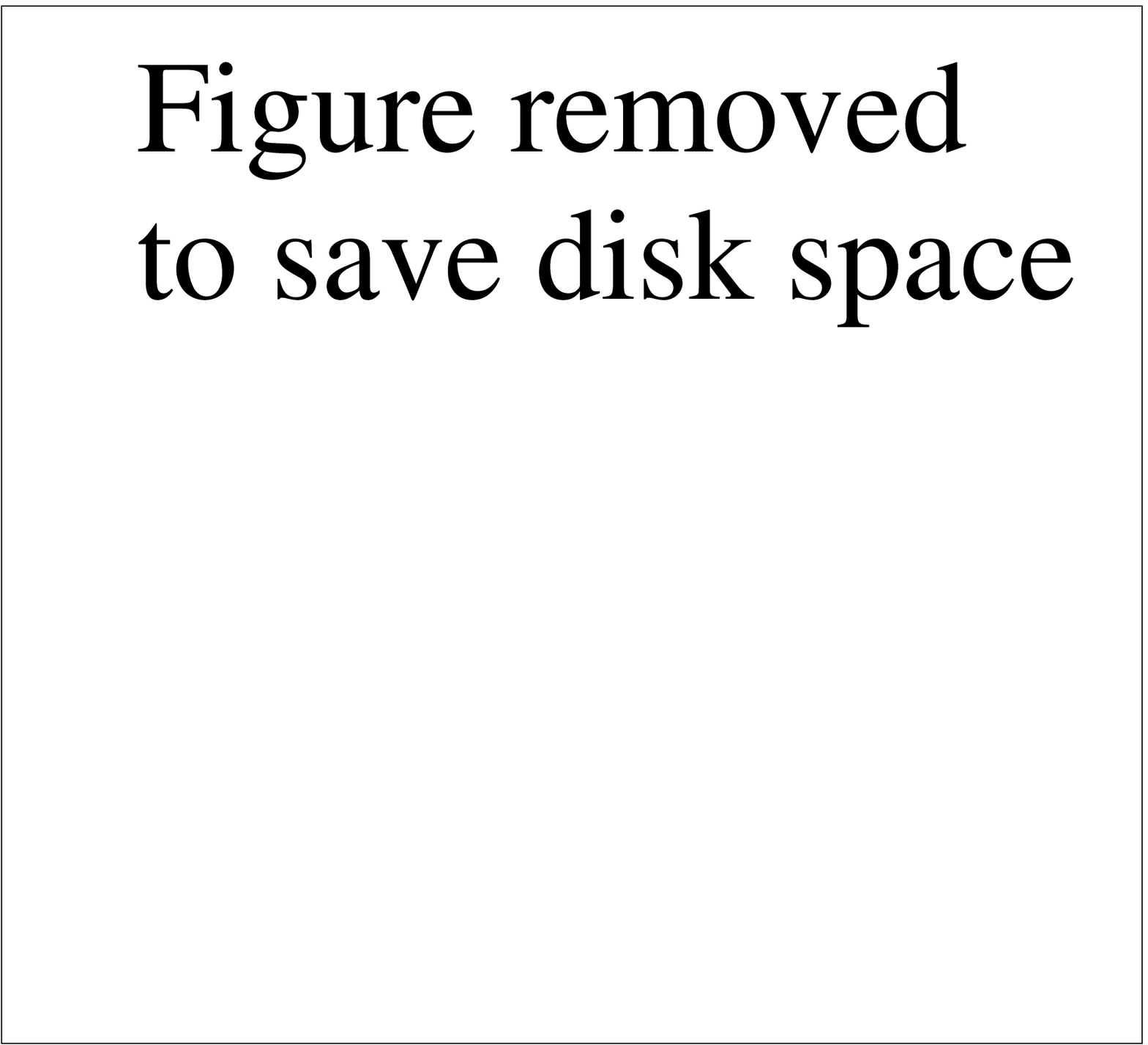}{0.0in}{0}{300.}{265.}{50.}{0.}
\caption{Top - The bulk of ACIS-I image of RCW 38 at low resolution.
Bottom - Matching field in a 2MASS K-band image.
The field of view in these figures is 16\arcmin $\times$ 16\arcmin.
Regions of interest are identified. The small ovals near the core are the ``common background regions'' used for sources in the core as discussed in \S 2.2. At 1.7 kpc 121\arcsec = 1~parsec. 
}
\label{acisfig}
\end{figure}

Time-dependent gain corrections were applied and acis\_process\_events rerun. While there were over 3$\times10^6$ level 1 events, our ``cleaned'' data file used for analysis contained 171,973 events. The central portion of these data is shown in Figure~\ref{acisfig}.

An exposure map was created using {\it merge\_all}, accurately representing the effective exposure time for a 1.7 keV photon. We chose this single energy for the exposure map since it is intermediate between the maximum of the effective area of the HRMA/ACIS system and an estimated mean source energy of $\sim$ 2.0 keV.  The exposure map corrects for the changes in the effective area of the mirror as a function of off-axis distance and telescope dithering.  The exposure map is later applied automatically by CIAO tools extracting count rates and spectra.

\subsection{Source Detection}
    PWDetect\footnote{http://cxc.harvard.edu/cont-soft/software/pwdetect.1.0.html}
     was used for source detection on our cleaned events list to identify sources across the entire I array. Threshold significance was set to detect sources between 4 and 5.31 equivalent Gaussian sigmas and the data are searched on scales of 0.5 to 16\arcsec.  With these settings, a false detection rate of 
$<1\%$ is expected. We did not change any PWDetect settings.  
Source detection was done on the whole field at once. The program did not seem to be confused by faint sources embedded in diffuse emission. 

   PWDetect produces a regions file defining the source extent for extracting photons.  These regions were adapted to more meaningful regions based on the actual $Chandra$ PSF and chip position for each source.  {\it mk\_psf} was used to obtain images of the PSF at various off-axis angles $\theta$~(arcmin), and rotation angles $\phi$ (degrees), around the ACIS array.  At each source location an ellipse was generated to enclose 95\% of the total X-ray energy.  
This lead to the following derived parameterization applied to our sources for elliptical regions around PWDetect's returned source locations:
$$
major~~axis =1.97-0.22\theta+0.15\theta^2 \\
$$
$$
minor~~axis =2.03-0.13\theta+0.08\theta^2 \\
$$
$$
angle =95.4+0.47\phi
$$
This formulation was used, except as noted below, to define the extraction region for each source.

\begin{figure}[t]
\plotone{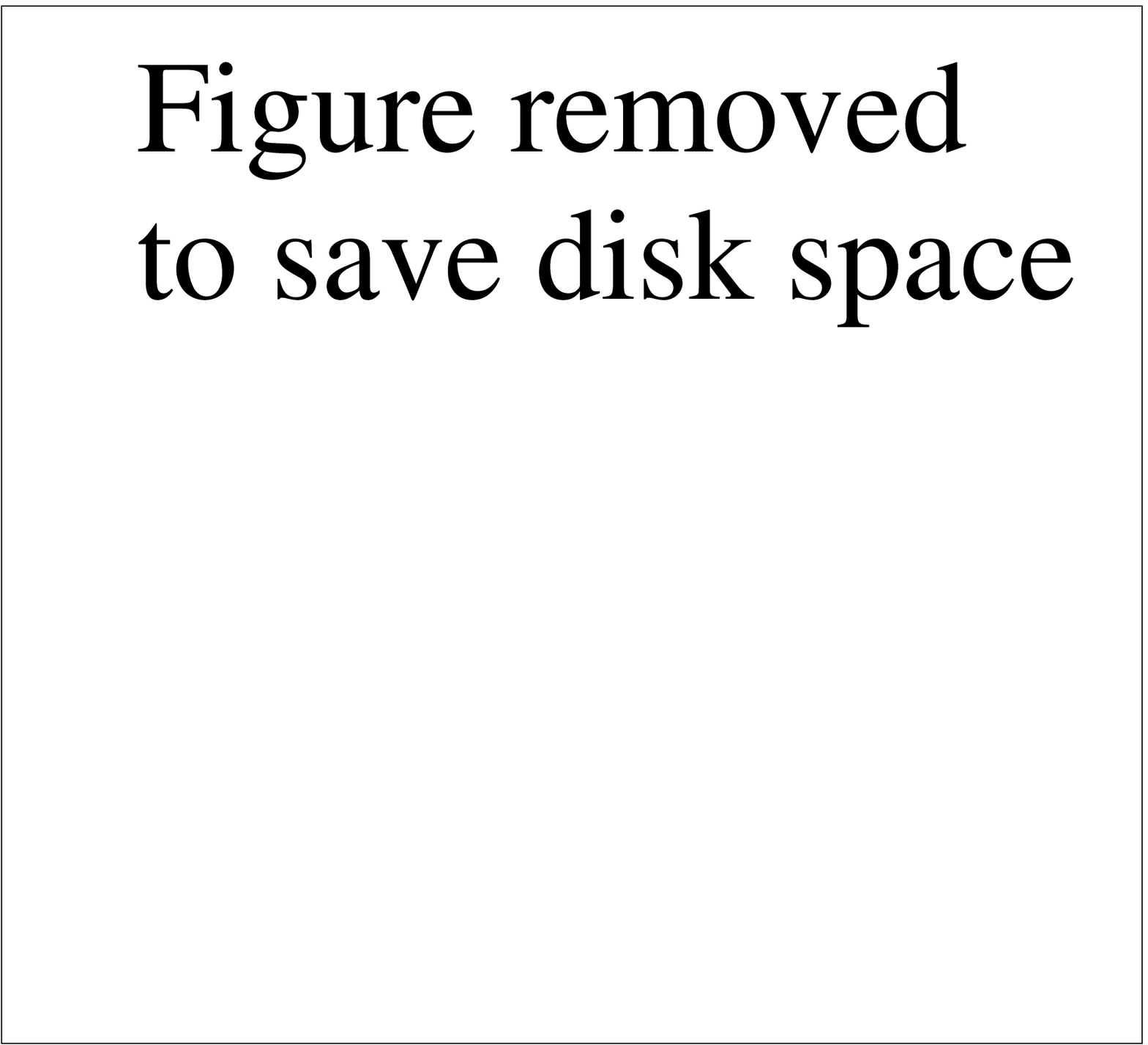}
\caption{VLT ISAAC K$_s$ band image of the central 2\farcm5 $\times$ 2\farcm5 region of RCW~38.  The image uses square--root scaling.
Contours of 0.5, 1 and 4 X-ray counts/pixel are overlaid.}
\label{sourcescentral}
\end{figure}

    Several spurious and overlapping sources were visually edited. To aid in editing unresolved and confused sources a near-infrared K$_S$ image with X-ray contours over-plotted proved helpful (Figure~\ref{sourcescentral}). The near-infrared data are discussed in detail in \S 6.

In most cases, background subtraction was made via background regions centered on each source. The background regions were elliptical annuli centered on the source with inner axes 2.7 times the length of the axes for source region and the outer axes 4.1 times the source region axes.  The inner edge of this annulus was chosen to make certain no energy from the extended wings was included in the background. The outer axis of the background annulus was chosen to be 4.1 times the source axis to allow the maximum background area which would not generally need to be adjusted for other sources. Background regions were manually adjusted in cases where chip edges or other sources interfered.  For areas near the center of the cluster, with several nearby sources, a common background region was created from a suitable region nearby.  
Sources in the central arcminute of the cluster were divided into two sectors, northwest and southeast. In each half a common two part background was used since the diffuse emission is present in the region which changes spectrally from northwest to southeast. The two background regions are indicated by the ellipses in Figure~\ref{acisfig}.

A total of 460 sources were found on the ACIS-I array, including 31 sources with more than 200 net counts, 49 sources with 100-200 net counts, 
71 sources with 50-100 net counts, and 78 sources with 20-50 net counts.  
Sources 95 and 228 were removed from the list as duplicates late in the analysis process, hence they are skipped in the enumeration. All X-ray sources are listed in Table~\ref{Sources}. The first column of Table~\ref{Sources} gives the internal identification of each source while column 2 gives the official IAU/CXC designation.  Columns 3 and 4 list the right ascension and declination (J2000).  Columns 5 and 6 list the off--axis distance in arcminutes and the net counts background subtracted and corrected for the aperture size respectively while columns 7 and 8 lists the number of intervals of constant flux at 95\% and 99\% significance requirements.  While details of this last column are discussed in \S 4, the key point is that more than one interval indicates a variable source.

\section{Cluster Membership} 
As stars progress from Class 0 to Class III (Lada \& Shu 1990), it is only with an unbiased sample of cluster membership that the evolution of the star+disk system can be understood.  Since optical spectroscopic follow up is not possible for such an absorbed/obscured cluster we develop here a method of using the 
X--ray color and spatial properties of each source to assess the probability that it is a cluster member independent of its infrared properties.
The typical method of model independent spectral analysis is to use X-ray colors in the form of hardness ratios, HR=(Cts$_h$-Cts$_s$)/(Cts$_h$+Cts$_s$) where ``h'' and ``s'' refer to the hard and soft bands respectively. 
Hardness ratios are of inherently limited value because of biases in the selection of bandpasses which lead to very non-uniform errors and limited dynamic range.  Sources with low counts tend to be driven toward the center of the distribution and there is a fundamental difficulty in breaking the degeneracy between temperature and absorption.  Instead, we use a quartile analysis technique for model free analysis of X-ray data explored by Hong \e 
(2004). Quartile analysis avoids some biases inherent in the selection of bandpasses needed to calculate hardness ratios. In this form of quartile analysis, one starts with the full ACIS band pass of E$_{lo}$= 0.3 keV and E$_{up}$ =8.0 keV.  $E_{x\%}$ is defined as the photon energy below which 
x\% of the photon counts are found and $Q_x = {(E_{x\%}-E_{lo})\over(E_{up}-
E_{lo})}$ is defined to be the normalized quartile.
The ratio of the bottom to top quartile (x= 25 and 75respectively) is representative of a two-point slope of the spectrum. For the case of a single temperature, the median, $m$($=Q_{50}$), is a function of the absorption. The quartiles are not independent, as the absorption changes the quartile ratio for a given temperature. Hong \e (2004) plot the data by normalizing the quartile ratio axis as $3\times Q_{25}/Q_{75}$ while the median is compressed as $\log (m/(1-m))$. On such a plot one can distinguish changes in temperature from extinction and can even distinguish thermal and non-thermal changes.

\begin{figure}[t]
\plotfiddle{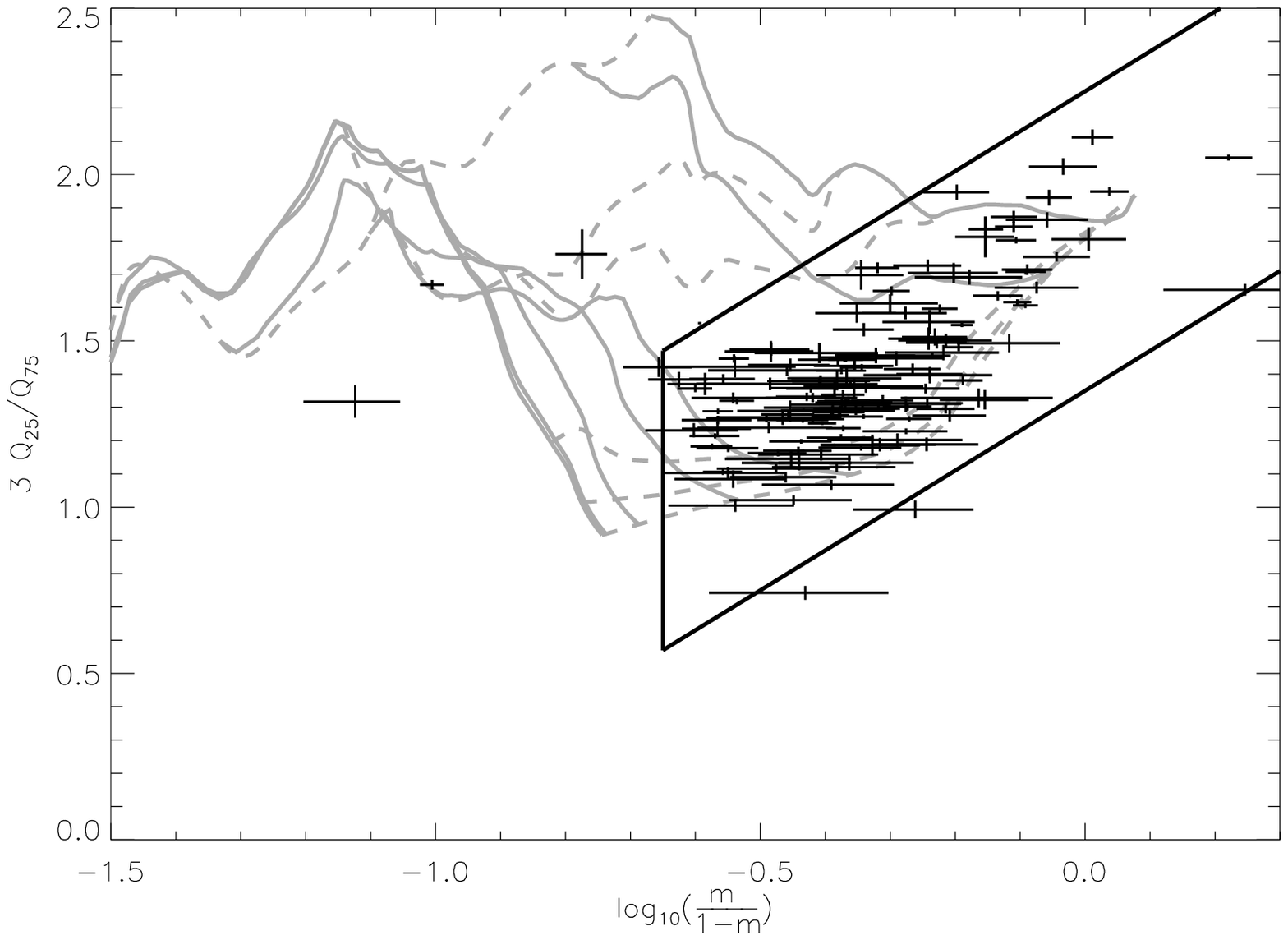}{0.0in}{00.}{300.}{265.}{50.}{-1.0in}
\plotfiddle{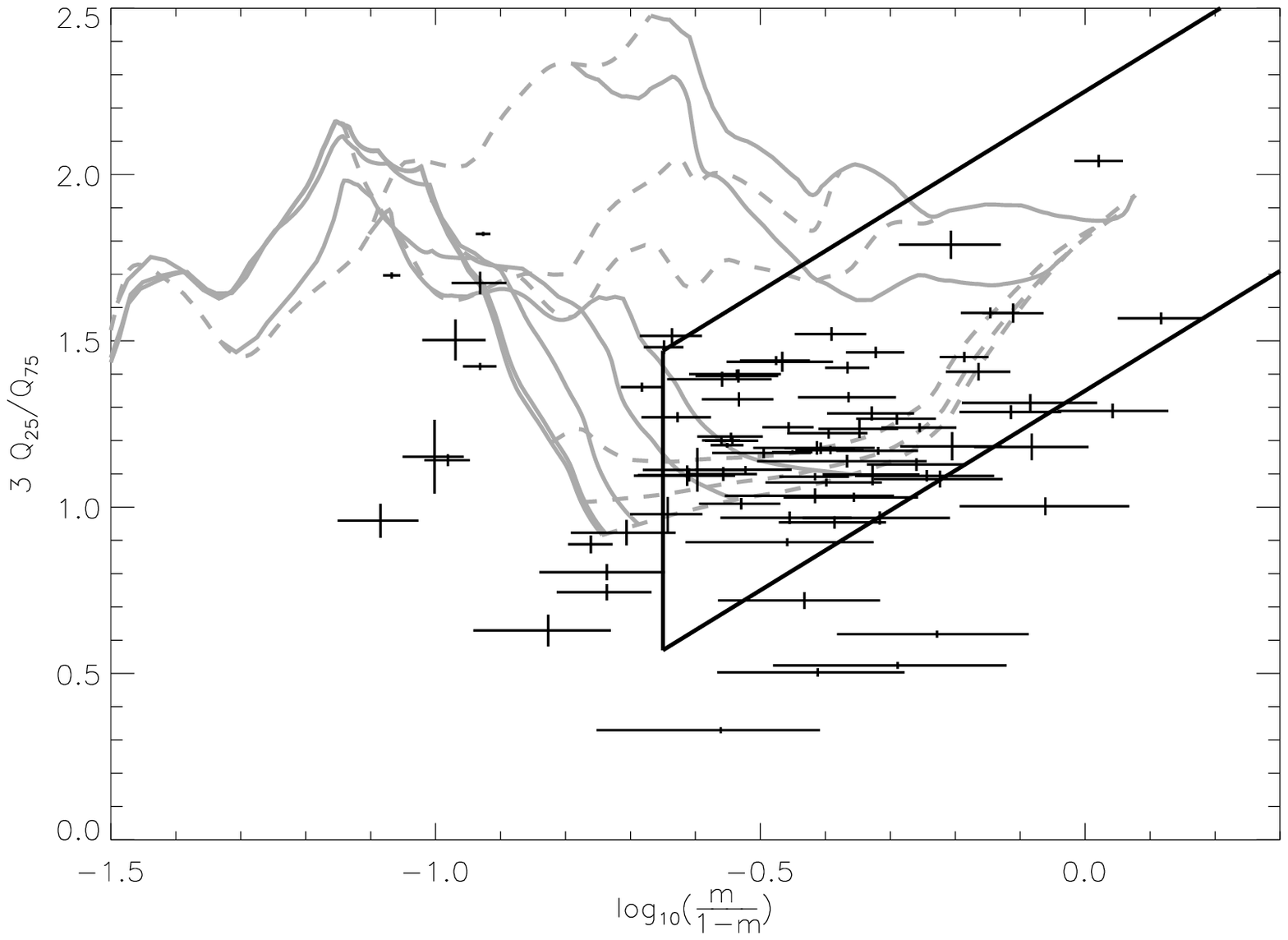}{0.0in}{00.}{300.}{265.}{50.}{-1.0in}
\caption{
Top: Plot of the normalized quartile values for stars within 200\arcsec\ of the cluster center.  A grid of temperature and absorption is overlaid following Hong \e (2004).  The solid lines are lines of constant \nh\ 0.1, 0.5, 1.0, 5.0 and 10 $\times 10^{22}$ from bottom to top.  The dashed lines are lines of constant temperature kT= 0.1, 0.5, 1.0, 5.0, 10 and 30 keV from left to right.  Errors are plotted. 
The parallelogram outlines the region occupied by probable cluster members. The diagonal lines are offset $\pm$ three deviations from a fit to all the bright sources to the right of the vertical demarcation. See text for details.
Bottom: Same as top, but for sources more than 200\arcsec\ of the cluster center.  For clarity, only stars with more than 30 counts are plotted.} 
\label{quant1}
\end{figure}

To determine if cluster members occupy a common region in such a quartile diagram we study the most probable cluster members, those source that lie within 200\arcsec\ of IRS~2. By calculating the number of X-ray sources within 1\arcmin\ of each X-ray source we find the distribution of X-ray sources is sharply peaked with a full width at half maximum of about 200\arcsec.
Plotting all such X-ray sources in the diagram shown at the top of 
Figure~\ref{quant1} shows there is a clustering of the data in color space. The bottom panel of Figure~\ref{quant1} shows the same axes but now for sources {\em more} than 200\arcsec off axis.  The difference is striking. Of the 63 sources with over 100 counts and less than 200\arcsec\ off-axis, 62 lie between
$-0.65 < \log(m/1-m)) < 0.25. \log(m/1-m)) =-0.65$ is the vertical line in
Figure~\ref{quant1}.   We fitted the distribution of the data between those two points using a two-variable linear regression first order equation 3$\times Q_{25}/Q_{75}=1.2 \times \log(m/(1-m))$+1.8 with a mean absolute deviation (MAD; Beers \e 1990) of 0.15.  The diagonal lines in 
Figure~\ref{quant1} are 3 deviations above and below this fit. Sixty-one of the bright sources near the cluster center could be found within three deviations of the fit. One point was clearly not a cluster member (Source 39) and the other point was on the fringe. This is indicative of a group of sources with similar extinctions and spectral signatures.
  
There are 209 X-rays sources within 200\arcsec\ of the IRS~2 and with less than 100 counts.  Of these, 36 have between 50 and 100 counts and all lie within 3 
MAD of the best fit line. An additional 54 have between 30 and 50 counts.  Two of these (Source 34 and Source 175) are clearly too unabsorbed (too low a value in the X-axis) to be considered cluster members.  Three others lie just below the -3$\sigma$ deviation demarcation but we {\em do not exclude} these are candidate members since the measurement errors can account for this small difference. The remaining 119 sources have less than 30 counts, four of these are discrepant by having too low of value along the X-axis and we exclude these sources from cluster membership.  Fifteen lie just below the -3$\sigma$ deviation demarcation.  This is more than twice the rate of such sources seen with 30-50 counts. Again we do not exclude these sources, we instead conclude that the measurement errors associated with the quartiles is biased such that sources with low counts have systematically lower values of $Q_{25}/Q_{75}$ and that the errors are twice as high for sources with least than 30 counts compare to those with 30-50 counts. This last point is consistent with the results from Hong \e (2004). Overall we exclude seven X-ray sources within 200\arcsec\ of IRS~2 from cluster membership based on the quartile analysis.  We consider the 202 remaining X-ray sources to be probable cluster members. 

There are 183 X-ray sources more than 200\arcsec\ from IRS~2. One hundred and ten of these lie within the $\pm 3\sigma$ deviation limits described above.  We chose these limits to be somewhat inclusive and to error on the side of membership.  We exclude from probable cluster membership 94 sources which are outside of these limits and more than 200\arcsec\ from IRS~2. Along with the seven sources nearer to IRS~2, the identification of 101 sources in the field as background (or foreground) field objects is consistent with expectations derived from the Champlane survey (Hong \e 2005). Our observations reached a mean flux limit of about $2 \times 10^{-14}~ \erg\ \ps \psqcm$ (2-10 keV) averaged across the field. Preliminary results from the Champlane survey for $Chandra$ fields in the plane of the Galaxy away from the galactic core indicate that we can expect to find about 70 background AGN, cataclysmic variables, neutron stars, black holes, and other non--PMS star point sources in the ACIS-I field.
Some examples include sources 14, 18 and 113 which are quite hard and absorbed in X--rays and yet appear unabsorbed at near-IR wavelengths (see \S 6).
The Champlane effort excludes soft X-ray sources such as dMe stars and white dwarfs and thus should be taken as a lower limit. 

The probable cluster members are noted in Table~\ref{quantin}. Probable non-members are noted in Table~\ref{quantout}. In these tables the first column is the source number, followed by the quartile values, $Q_{25}$, $m$ and $Q_{75}$, with errors interleaved.  The plot values are calculated for convenience and listed in the final two columns. 

\subsection{Cluster Density}
\begin{figure}[t]
\plotfiddle{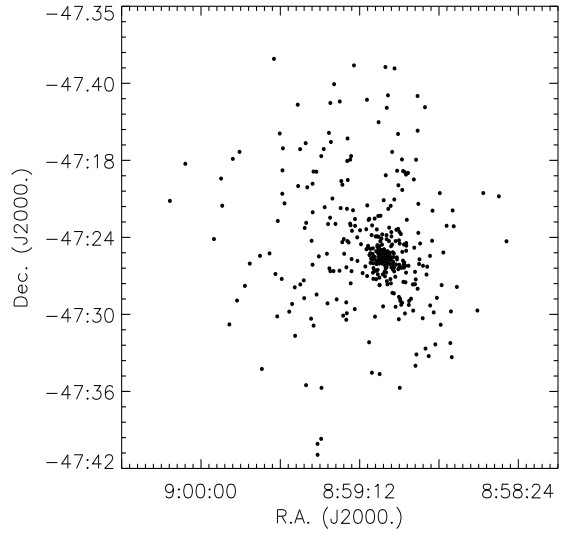}{0.0in}{90.}{250.}{320.}{100.}{-1.0in}
\plotfiddle{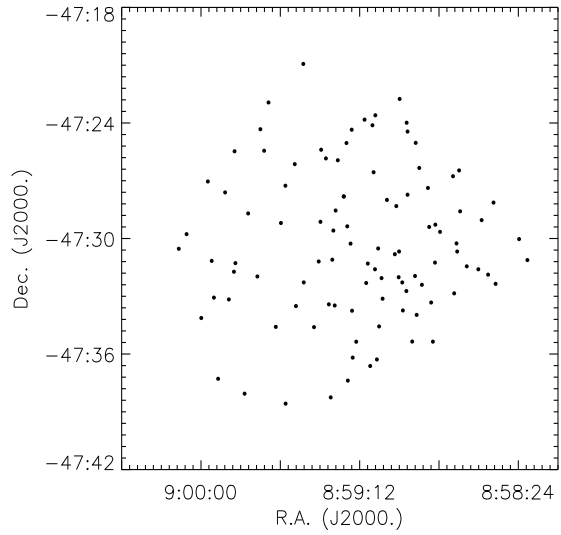}{0.0in}{90.}{250.}{320.}{100.}{-1.0in}
\caption{
Top: Position plot of all probable X-ray cluster members as determined by quartile analysis.  Bottom: Position of non--members. A density analysis shows that this distribution is uniform.}
\label{posplot}
\end{figure}

We plot the positions of the members and non--members in Figure~\ref{posplot}.  It is clear that the members are clustered while the non--members have a more uniform distribution.
A density analysis was performed by calculating the number of cluster members within 15\arcsec\ of each cluster member.  
Asymmetries in a plot cluster density as a function of the distance from the dominant ionizing source IRS~2 indicate that it is not the center of the cluster. The geometric centroid was calculated in terms of stellar (X--ray source) density, {\em not} mass density, using cluster members within 100\arcsec\ of IRS~2, and is located at: 08h59d04.64s, $-47^{\rm o}30\arcmin44\farcs00$.  This is displaced 10.4\arcsec\ to the west southwest of IRS~2 and aligned roughly with the 10~\micron\ dust ridge reported by Smith \e 
(1999). Varying the inclusion radius by 50\arcsec\ can change this centroid location by 6\arcsec, this is mostly controlled by the initial guess that the center was near IRS~2.

\begin{figure}[h]
\plotfiddle{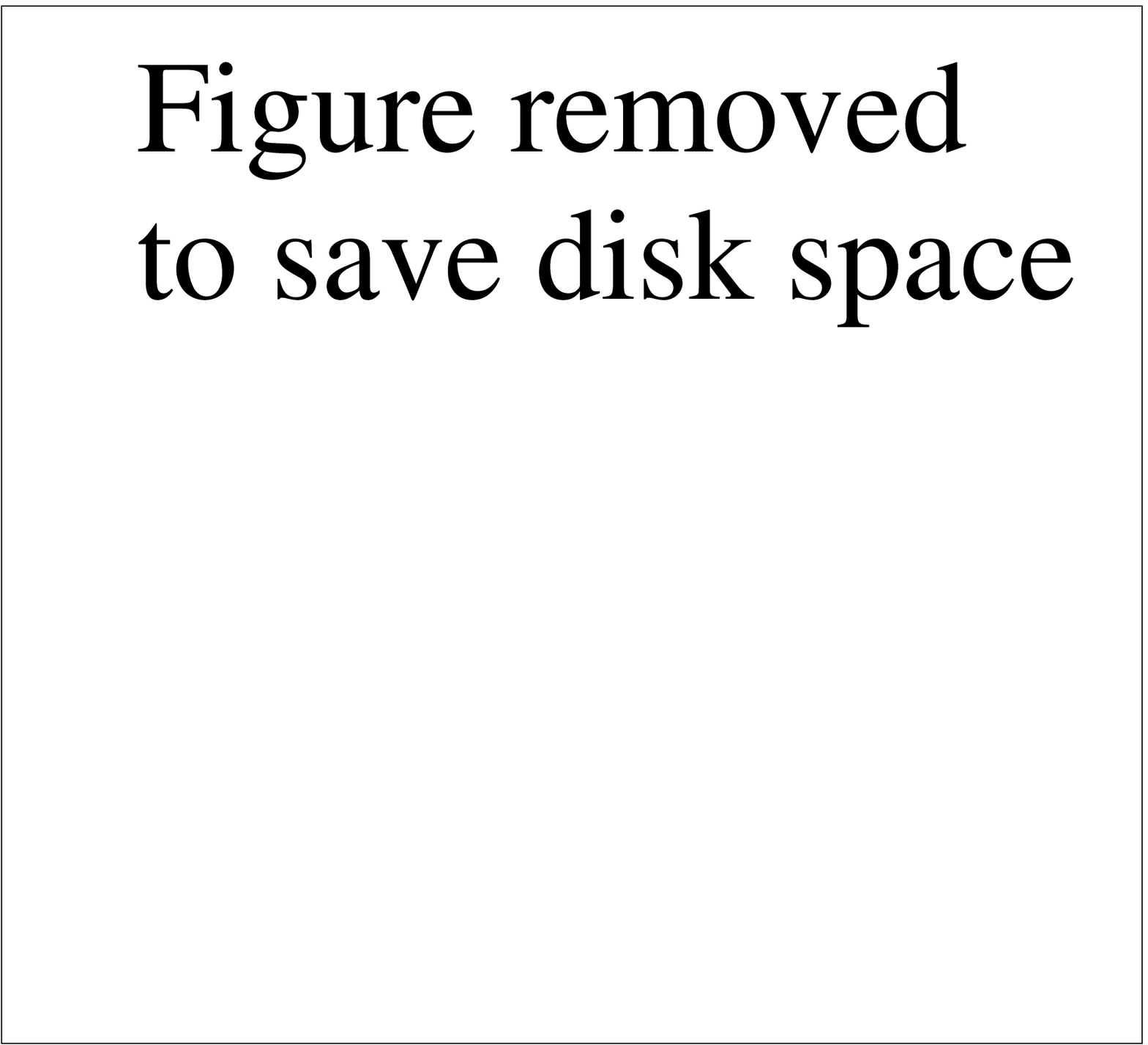}{0.0in}{90.}{250.}{320.}{50.}{-1.0in}
\plotfiddle{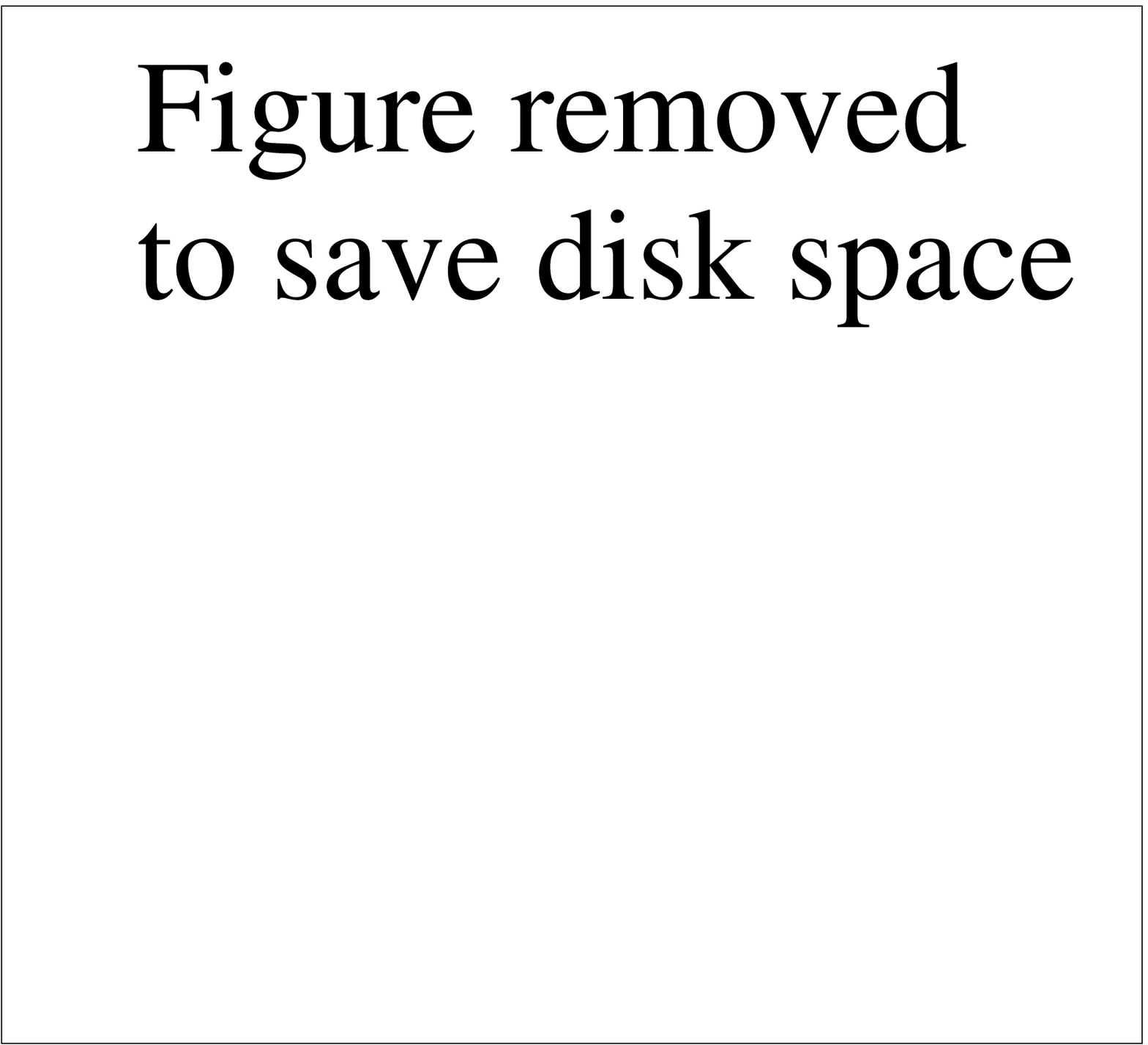}{-1.0in}{90.}{230.}{300.}{50.}{-1.0in}
\caption{
The X--ray source density profile of the RCW 38 cluster. 
Top: For each probable cluster member the distance from the cluster center and the number of cluster sources within 0.25 pc (15\arcsec) are plotted.
The line is fitted to the data. 
Bottom: A similar plot for the probable non-members. The bimodal behavior is due to the results being scaled up from 15\arcsec\ samples.}
\label{cluster_dens}
\end{figure}

The results of the full density analysis are shown in Figure~\ref{cluster_dens}.   
The cluster density drops to half the peak density of about 100 X-ray sources per arcminute at about 15\arcsec\ ($\sim$0.12 pc) from the cluster center.
We find the cluster is very sharply peaked with broad wings relative to a 
Gaussian profile -- hence, this is not a relaxed system.
We fitted the data in Figure~\ref{cluster_dens} as an exponential of the number of sources per parsec$^2$ $\sim exp(-5.0 d)$ where $d$ is the distance from the cluster center in parsec.  The peak cluster density, as measured via this fit, is about 400$\pm 20$ X-ray sources per parsec$^2$ 
(=100$\pm 10$ X-ray sources per square arcminute).  The density of the non-members is 2.3 per square arcminute $\pm 1.1$.  
A plot of the density profile of these sources shows a slight enhancement about 
100\arcsec\ from the cluster centroid indicating that as many as 8 of the candidate non--members have been misidentified as such.  While the quartile system as applied here is not a perfect system for determining cluster membership based on X-ray colors, it appears highly effective.

We can use these data to estimate the total cluster size.  Recent work on mass stratification in the ONC by Feigelson \e (2005) indicates that at our sensitivity limit of log L$_x \sim 30.0$ one detects about 16 percent of the stars in a cluster that is similar to the ONC.  Comparisons to measurements of the ONC are quite appropriate to RCW 38 as both clusters have mid to early O stars. Since our sensitivity was a little below this limit we expect that we are detecting at most 15\% of the cluster members. This should include half of all stars above 0.5 M$_\odot$.
Scaling from the 360 members detected we estimate that the total cluster size could be as high as 2400 members with a peak density of about 1000 sources in the central parsec$^2$.

\section{Variability}

In this section, we examine the variability of the X-ray source population.  Variability studies allow us to begin to assess the plausibility of X-ray generation mechanisms.  These can be constrained by the timescales and flux changes observed in the variability. Variability also offers corroborating evidence that a source is stellar in nature and hence a possible cluster member.  Unlike other X-ray sources almost all coronal sources vary given a long enough observation (see Getman \e 2005).  These variations can either be stochastic or impulsive deviations from constancy.  

Various methods can be employed to investigate variability. This topic has been reviewed briefly by Wolk \e (2005) and more thoroughly by G\"udel (2004) 
and Favata \& Micela (2003). One rigorous technique commonly used is a 
one--sample Kolmogorov-Smirnov (KS) test to identify if the photon arrival times are consistent with a constant rate.  The KS test does not give any information on the nature of the variability in objects.  We identify X-ray sources as variable if they have a KS statistic less than 0.001.  That is, the cumulative event arrival time distribution deviates from a linear distribution at a confidence greater than 99.9\%.   Amongst the brightest 31 sources (those with over 200 net counts), 7 (23\%) were variable at $>$ 99.9\% confidence.   
That rate is maintained among the 80 sources with over 100 net counts, of which 
19 (23\%) were variable at $>$ 99.9\% confidence.  

\subsection{Flaring} 
About 25\% of the bright sources are detected as variable in 100ks, but the KS test tells us nothing about flaring. As demonstrated in
Wolk \e (2005), the use of Bayesian Blocks (BBs; Scargle 1998) provides a method of flare detection without the biases inherent in binning the data.
In some senses the BB method is similar to the KS test; the existence of more than one block indicates that the flux has changed at a certain confidence level. However, while the KS test reports very little about the nature of the change, the number of BBs and measurements of the mean rate, $\sigma$ and duration of those blocks allow quantitative analysis.
The BB method is explicitly designed to avoid binning the observation into equally spaced time intervals.  Bayesian Blocks conform to complex X-ray 
lightcurves remarkably well (see Getman \e 2005).  We tested each lightcurve with a ``prior ratios'' set to approximate both 95\% and 99\% confidence that a flux change has occurred. 

Using the BB technique, 7 of the 31 sources (22.5\%) with over 200 counts and 21 of the brightest 80 (26\%) could not be fitted with one block at a constant level with over 99\% confidence. This number is very similar to the 23\% which varied at similar confidence as measured by the KS test.  Overall 37 sources required more than one BB.  These results are tabulated as part of
Table~\ref{Sources}.

The BB method converts the lightcurve to temporal periods of constant flux, thus, one can measure the amount of rise between the blocks and estimate the rate of change between blocks.  In their study of the extremely deep COUP data set, Wolk \e (2005) find that most stellar sources have a characteristic rate, R$_{char}$ and found sources were at their characteristic levels for about  75\% of the time.  They further found that a normalized rate of rise  $\Delta
\equiv 1/R_{char}\times dR/dt > 10^{-4} $s$^{-1}$ was indicative of a flare.   
Following Wolk \e we define dR as the difference between the rates of adjacent blocks and dt as the shorter of the two blocks. We choose the value of the minimum block as the characteristic rate since we do not have the long observation time to define a true characteristic level.  

\begin{figure}[h]
\epsscale{0.8}
\plotone{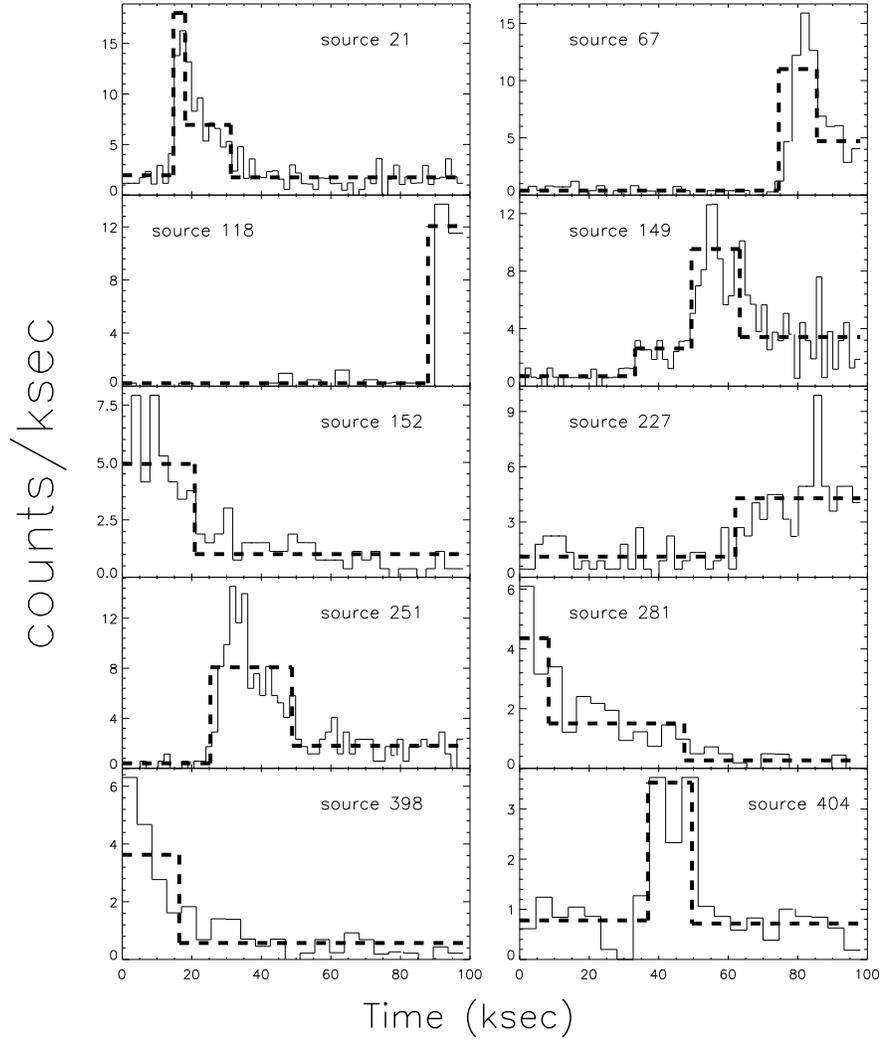}
\caption{
Flares on bright stars. The Figures show lightcurves of all sources with over 100 counts and flares (21, 67, 118, 149, 152, 227, 251, 281, 398 and 404).  All are cluster members. Sources 149 and 251 are among the hardest X-ray sources, Source 251 is a possible embedded massive star. Source 118 is also deeply embedded (see text). Histogram bins traced with a thin solid line are to guide the eye (Bin sizes were selected to yield an average of 10 counts per bin). Blocks of constant flux (99\% confidence according to Bayesian statistics) are traced in by the thick-dashed line. }
\label{bflares} 
\end{figure}

\begin{figure}[h]
\epsscale{0.8}
\plotone{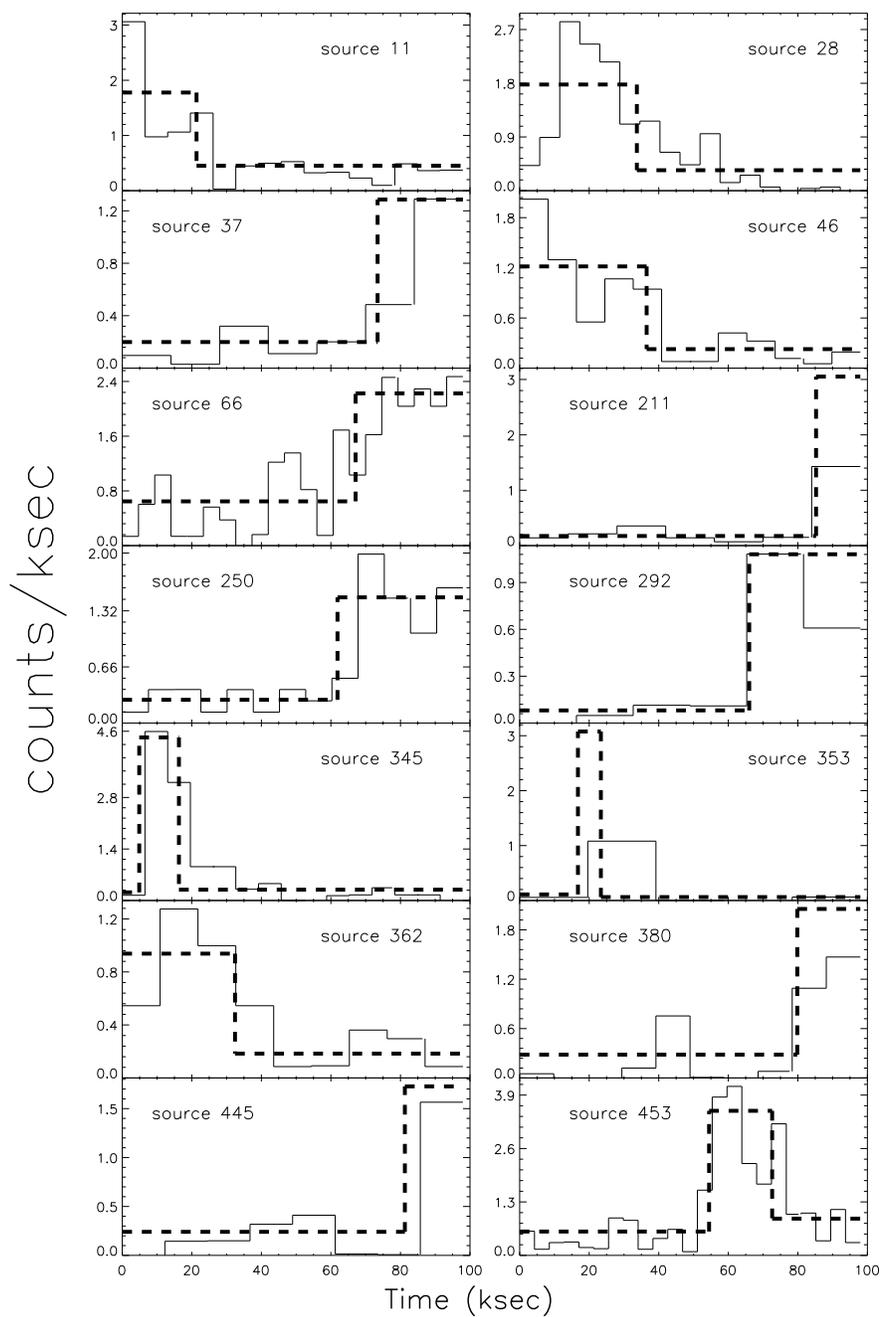}
\caption{
Flares on fainter sources.  Lightcurves of 14 X-rays sources with flares and $<$ 100 total counts. Sources: 11, 28, 37, 46, 66, 211, 250, 292, 345, 353, 362, 380, 445 and 453.}
\label{modflares} 
\end{figure}

\begin{figure}[h]
\epsscale{0.8}
\plotone{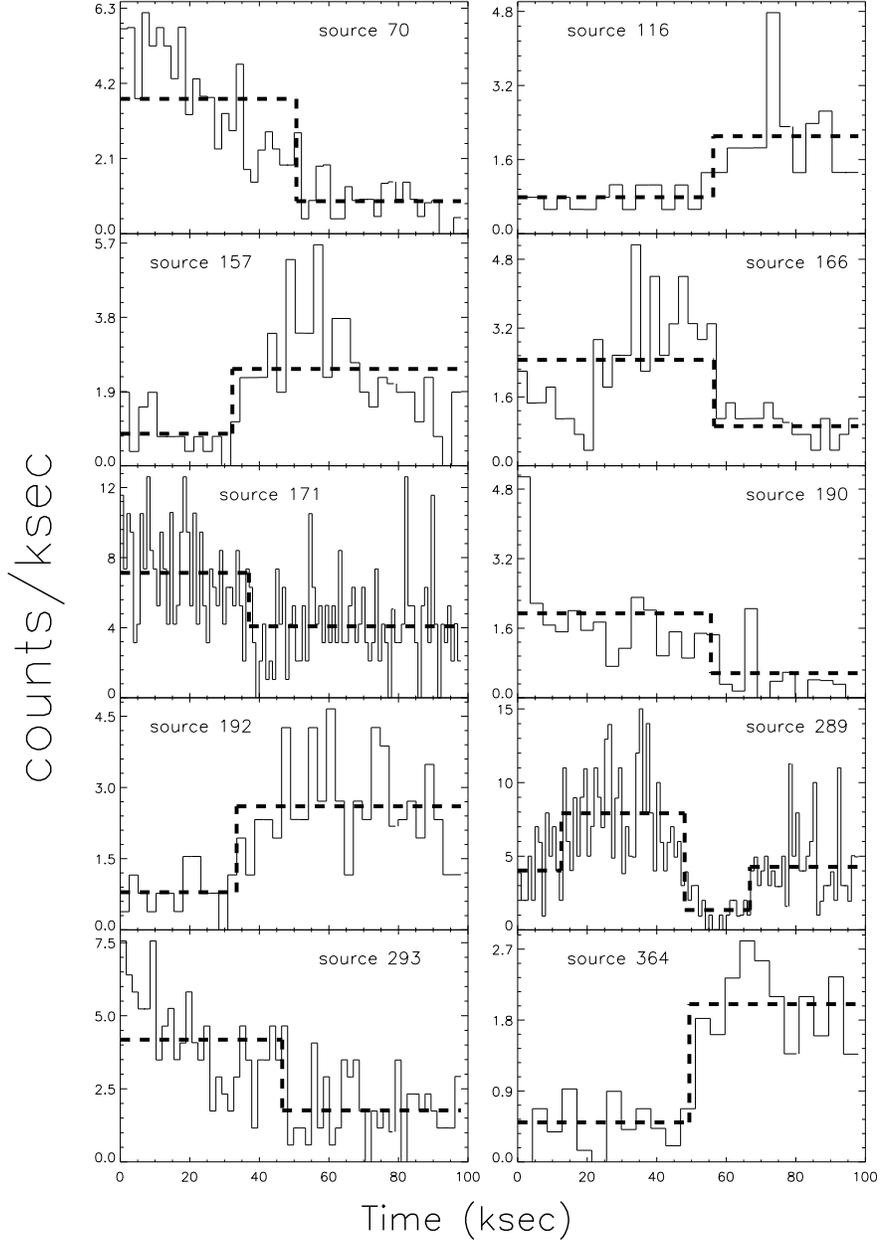}
\caption{
Lightcurves of the 10 X-rays sources seen to vary at 99.9\% confidence and which did not flare based on the criteria given in Wolk \e (2005). 
All of these sources are probable cluster members.
Source 289 is the only source in the cluster to show 4 distinct flux levels at 99.9\% confidence.}
\label{noflares} 
\end{figure}

Twenty-four sources have this form of impulsive variability (see 
Figures~\ref{bflares}, \ref{modflares}).  
Wolk \e (2005) further use a criterion that each flare be bright. To do this they use the error on the BB count rate as $\sigma$ and define ``bright'' in the sense that the rate in the flaring block is 2.5 $\sigma$ brighter than the characteristic block. $$R_{flare}- 2.5\sigma > R_{char}*1.2$$  
All 24 flares met this criterion.  No star was seen to flare twice.
Figure~\ref{noflares} shows lightcurves of ten sources which varied in
X-rays but were not seen to flare.

\subsection{Flare Rates}
We can use these data to assess flare rates.  Since the fraction of stars seen to vary (via either the Bayesian blocks of KS method) does not decrease as we limit the total counts from 200 to 100, we assume that the test is fully valid at this count rate.  There are 103 stars with over 75 counts, of which 23 vary (30\%). This fraction exceeds the variability rate seen at higher fluxes so we do not appear to be missing any true variable.  However, of 48 sources with between 50 and 75 counts, only 5 (10\%) vary.  Of the 87 sources with between 30 and 50 counts, 7 (8\%) vary.  Clearly there is a roll-off in the sensitivity to variability below 75 counts. To prevent biases in the data we constrain the following analysis to sources with over 100 counts. 

Nine of 72 (12.5\%) probable cluster members with over 100 counts flared in 96.7 ks. This represents a little less than half of the bright cluster members which are detected as variables.  Non-member Source~404 flared, none of the 6 other bright non-members varied.  Assuming that all stars are the same and that there are no stars more prone to flare than others then we conclude there are about 775~ks between flares.  This is in between the values obtained for Solar mass stars in the ONC (one flare per 640 ks; Wolk \e 2005) and all stars in the much older cluster -- NGC~2516 (About 1 per megasecond; Wolk \e 2004). 

Finally, variability analysis can be used to assess the veracity of the quartile approach to cluster member selection. We expect stars to vary more than other  X--ray sources. Of the 35 stars seen to vary with 99\% confidence, 34 of them are identified as cluster members by the quartile approach.  This gives us great confidence among the bright sources. Further, variability was detected on sources down to 30 counts.   In the field there are 202 cluster members with between 30 and 100 counts, 16.8\% of those vary.   In contrast, only one of the 
22 non-cluster members varies. The lone non-member to flare was Source~404.  
From spectral analysis we find that it is one of the most unobscured sources in the field and hence likely a foreground dMe star.

\section{Spectral Analysis}

     Analysis of the X-ray spectra of each source was performed to determine the bulk temperature of the corona and the intervening column of hydrogen. 
     Source and background pulse height distributions in the total band (0.3-8.0 keV) were constructed for each
     X-ray source.  Model fitting of spectra was performed using {\it Sherpa} (Freeman \e 2001).  The final fits were done with CIAO version 3.1.0.1. 
 The CIAO script {\it psextract} was used to extract source spectra and to create an Ancillary Response Function (ARF) and Redistribution Matrix Function (RMF) files. For sources with over 30 counts, data for each source were grouped into energy bins which required a minimum of 8 counts per bin and background subtracted. The optimization method set to Levenberg-Marquardt and $\chi$--Gehrels statistics were employed.  For sources with under 30 counts, models were fitted unbinned using “cstat” statistics (Cash 1979) and Powell optimization. 

\subsection{Choosing a Spectral Fitting Model}
For uniformity it is desirable to settle on a single model spectrum for all sources. 
To find the most appropriate model for our sources, we ran a series of test fits on the brightest 80 sources, those with over 100 counts.  
In these tests, the 80 sources were successively fitted with various absorbed thermal plasma models. These included the model from Raymond--Smith
(1977),  the ``mekal'' model  (Mewe \e 1985, Mewe \e 1986, Kaastra
1992, Liedahl \e 1995), APEC\footnote{An emission spectrum from collisionally-ionized diffuse gas calculated using the APEC code v1.10. More information can be found at http://hea-www.harvard.edu/APEC.}, thermal bremsstrahlung 
(Kellogg, Baldwin \& Koch 1975) plus blackbody and single power law models. The fitting was done using {\it Sherpa}, the absorption law was assumed to be
$abs(E) = e^{-\nh\ \sigma(E)}$  in which the photo-electric absorption cross-sections ($\sigma(E)$)  are taken from Morris \& McCammon
(1983). Absorption and temperature or the power law index ($\Gamma$) were left as fitted parameters.  A two temperature absorbed Raymond--Smith model was also tested.  To evaluate the results of the test, we used 
The mean was calculated after trimming away outliers more than 2 MADs from the median (similar results were obtained with a 3 MAD cut). The results are summarized in Table~\ref{test-res}.  In this table, the columns represent the various models, Raymond-Smith plasma, two--temperature Raymond-Smith plasma, APEC, mekal, thermal bremsstrahlung, blackbody and power--law.  The rows represent the mean $\chi^2$ per degree of freedom, the MAD and the number of outlying measurements rejected in the MAD determination process.  This last row indicates the stability of the model relative to the data. From the data in the table it is clear that the blackbody and power--law fits are poor for the bulk of the sources.  Among the single temperature fits, the Raymond-Smith and APEC fits are clearly superior to mekal and thermal bremsstrahlung for these data. The reduced $\chi^2$ tend to be slightly higher for both thermal bremsstrahlung and mekal, and the fraction of rejected fits is very high for the mekal models.

It was difficult to determine superiority between the APEC and
Raymond--Smith models.  APEC had a slightly lower mean $\chi^2$ per degree of freedom (d.o.f.), but more scatter than Raymond-Smith fits. 
APEC performed marginally better for the brightest sources, but Raymond-Smith performed better at the faint end.  We compared luminosities determined by the two models and found the standard deviation between the two models was about 20\%.  This difference is dominated by outliers, removing the 10 outliers lowered the deviation between the two models to 3\%.  When we compared the \nh\ fit to the outliers to the visual extinction (calculated in \S 6) we found that the \nh\ calculated using the Raymond--Smith model was more consistent with the independently derived A$_V$ and the independently derived \nh\ to A$_V$ relation.  For these reasons we chose the Raymond-Smith model for our main spectral fitting model.

 A few sources are fitted very well by a single power law, these are probably background AGN. Still, in most cases a thermal model (e.g. APEC or Raymond-Smith) is better even for non--cluster members. Similarly some very cool sources, probably white dwarfs, are best fitted as blackbodies (see \S 7.2).
Finally, some sources simply defy fitting by any one model.

\subsection{Global Fitting of X-ray Spectra}
When performing bulk processing on the catalog of $\approx 460$ X-ray sources we found it was advantageous to perform a preliminary fit to the data with a simple blackbody and use the result of this fit as the initial guess for the Raymond--Smith fit.  This ensured that the local minimum found by Levenberg-Marquardt optimization was realistic.  From the experiments above, we see that the two-temperature models were no better than the one-temperature fits.  This is expected in a high extinction environment as we have here. Any cool component which may exist is overwhelmed by soft absorption due to the large hydrogen column.  We attempted detailed two-temperature fits for all sources with over 100 counts. Of the 80 fits only 35 of them had more than 10\% of the flux contributed by each component and had temperatures of the two--components separated by more than 5\%. 
Two issues show up throughout the fitting process. 1) There is a fundamental degeneracy in the fits between \nh\ and kT, fits of similar quality can be obtained by increasing \nh\ and then correspondingly changing the temperature.  This is especially true in regions of high absorption such as RCW~38. 2) Since there is little sensitivity to X-rays above 10~keV, model fits are not very reliable much above this energy range. Because of this, if the fitted temperatures exceed 15 keV we simply label them ``$>15$'' in the tables. 

The fit results for the two--temperature fits are tabulated in Table~\ref{2t} including results for three non-cluster members.  In Table~\ref{2t}, Column 1 gives the source name from Table~\ref{Sources}, 
Column 2 gives the goodness of fit to an absorbed two--temperature Raymond--Smith plasma in terms of $\chi^2$/d.o.f. Columns 3--8 give the fit parameters \nh, kT$_1$,kT$_2$ and the associated errors. Column 9 lists the log of unabsorbed flux from 0.3-8.0 keV, while column 10 lists the ratio of the flux of the two components kT$_1$/kT$_2$.  
For the cluster members, the luminosity from 0.3-8.0 keV of each source is calculated based on the derived unabsorbed flux and the cluster distance of 1.7 kpc and listed in column 11. 
These are the brightest and best fitted cluster members and can be used to derive the basic cluster parameters.  The mean \nh\ of these stars is 2.6$\times 10^{22} $cm$^{-2}$ with a MAD of
0.2, a mean kT$_1 \sim 700$~eV (MAD =80~eV) and the mean kT$_2 \sim
4.7$~keV (MAD =190~eV). 
The mean kT$_1$ is very similar to the average value of about 800~eV seen in several other clusters (Sanz-Forcada \e 2003).

For the remaining cluster members we performed absorbed one--temperature fits using a two step fitting procedure. Initial conditions were set so that \nh=1.0 $\times 10^{21}$ cm$^{-2}$ and kT=1.0 keV. Then an initial fit was made with an absorbed thermal blackbody model.    
These fit results were then used as initial conditions for an 
absorbed Raymond--Smith plasma model. The results of the fits to the
Raymond--Smith models are tabulated in 
Table~\ref{1t}.  In Table~\ref{1t}, Column 1 gives the source name from
Table~\ref{Sources},  Column 2 gives the goodness of fit to an absorbed one--temperature Raymond--Smith plasma in terms of $\chi^2$/d.o.f. Columns 3--7 give the fit parameters \nh,
kT, associated errors and unabsorbed flux.  The luminosity of each source is calculated based on the derived unabsorbed flux and the cluster distance of 1.7 kpc and listed in column 8. The median \nh\ of these stars is 2.6$\times 10^{22} $cm$^{-2}$ while the median kT is $\sim 2.55$~keV.

 We separately tabulated fits for sources with less than 30 counts.
Cstat statistics and Powell optimization were used, appropriate for faint sources.  These fits have large, though non--systematic, errors (Table~\ref{1t-faint}).  The columns in this table are identical to the columns of Table~\ref{1t}.  The median \nh\ of these stars is 3.2$\times 10^{22} $cm$^{-2}$ while the median kT$_1$ is $ \sim 2.1$~keV.  Note that change to higher \nh\ would lead to a high kT as well {\it if the sources were similar.} It is easy to understand that the fainter sources are preferentially more absorbed, but it appears that they are also cooler.

Finally, we performed fits on the non-cluster members.  Only 29 of these have over 30 counts. Three were fitted as two-temperature plasmas, although the fits to sources 39 and 447 will be revisited later. 
Most of the 26 remaining sources were well fitted by the Raymond--Smith model with reduced $\chi^2 < 1.71$ for all and reduced $\chi^2$ between 1.3 and 0.3 for all but three others. 
The median kT for these sources was about 1.25 keV and the mean \nh\ was about 
$0.5\times 10^{21}$cm$^{-2}$. As a group, they are cooler and less absorbed than the cluster members. For completeness we tabulated fits for non cluster members with under 30 counts. These results are tabulated in Tables~\ref{non} and \ref{nonpoor}. The columns in this table are identical to the first 8 columns of 
Table~\ref{1t}. 
Thermal spectra of the brighter non--cluster members range from 140~eV to over 15~keV.  The softer sources may be foreground dMe stars or white dwarfs (see \S 7). In Table~\ref{nonpoor} the temperatures of 30\% of the sources exceed
15 keV (compared to 7.5\% in Table~\ref{1t-faint}), indicating these are probably power law sources and most likely background AGN.

\begin{figure}[t]
\plotfiddle{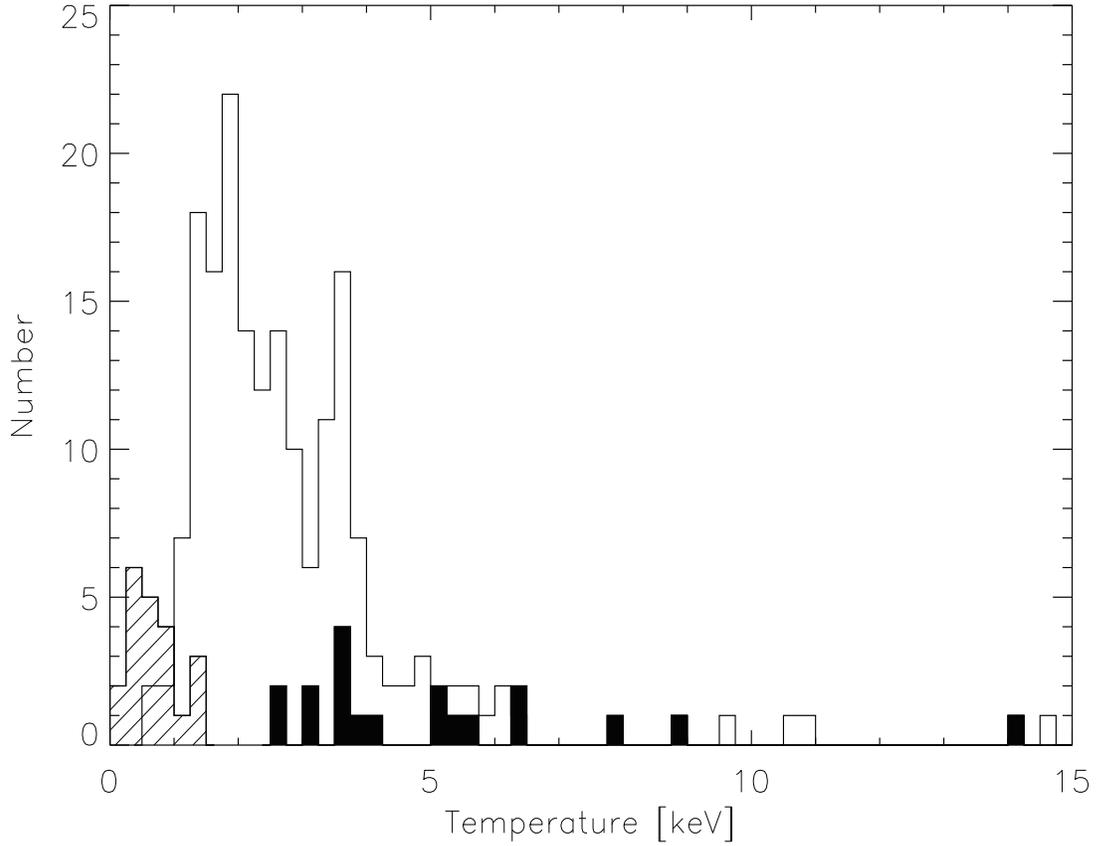}{0.0in}{90.}{400.}{520.}{-50.}{0.}
\caption{ Histogram of derived plasma temperatures for each cluster member. Fits for 188 sources fitted with one temperature Raymond--Smith plasma are indicated with the unshaded histogram.  The solid-filled histogram represents the high temperature component of the two--temperature fits.  The line-filled histogram represents the low temperature component from these fits. The bins are 0.25 keV wide.}
\label{kt}
\end{figure}

\begin{figure}[t]
\plotfiddle{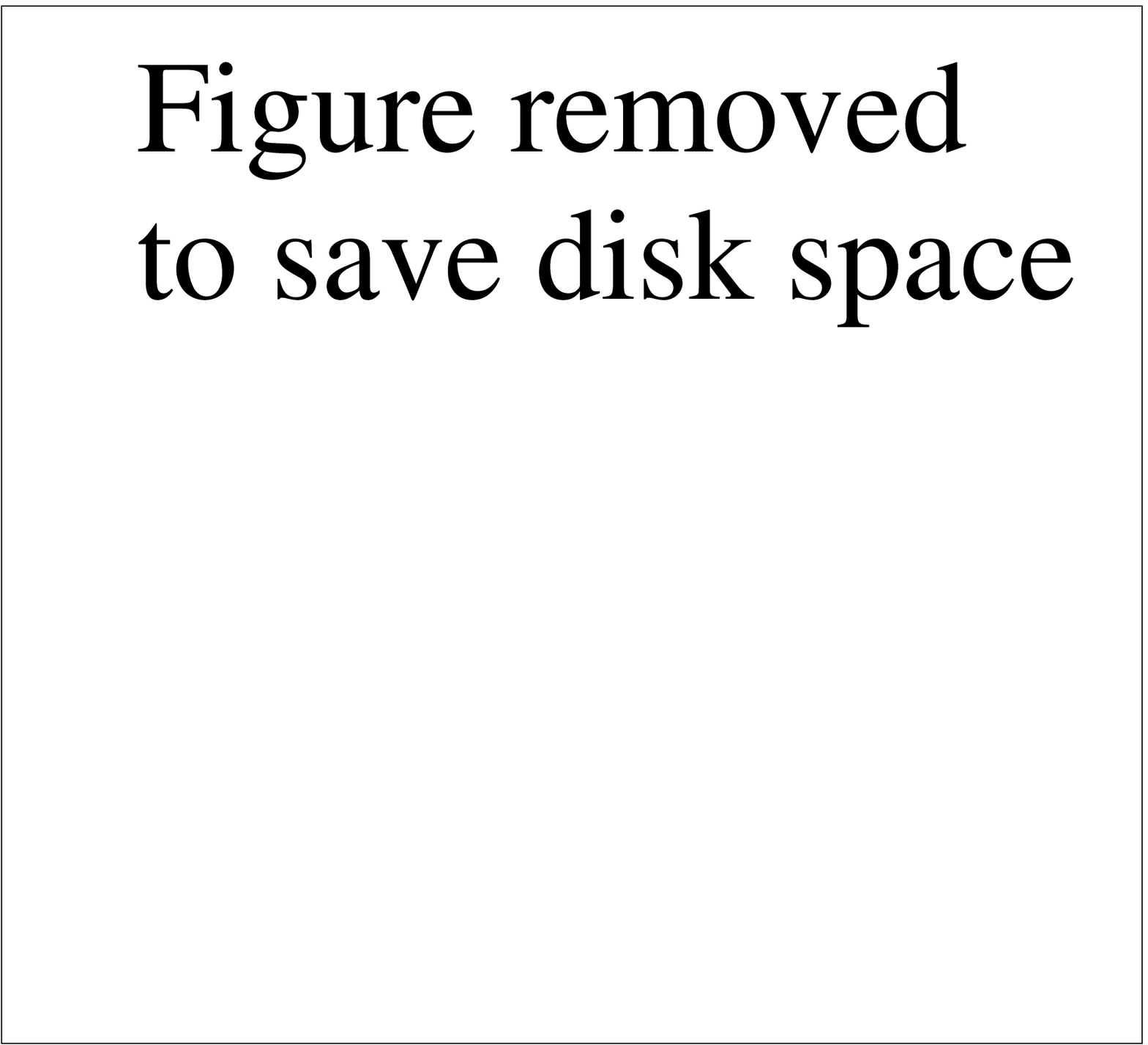}{0.0in}{90.}{200.}{320.}{50.}{0.}
\plotfiddle{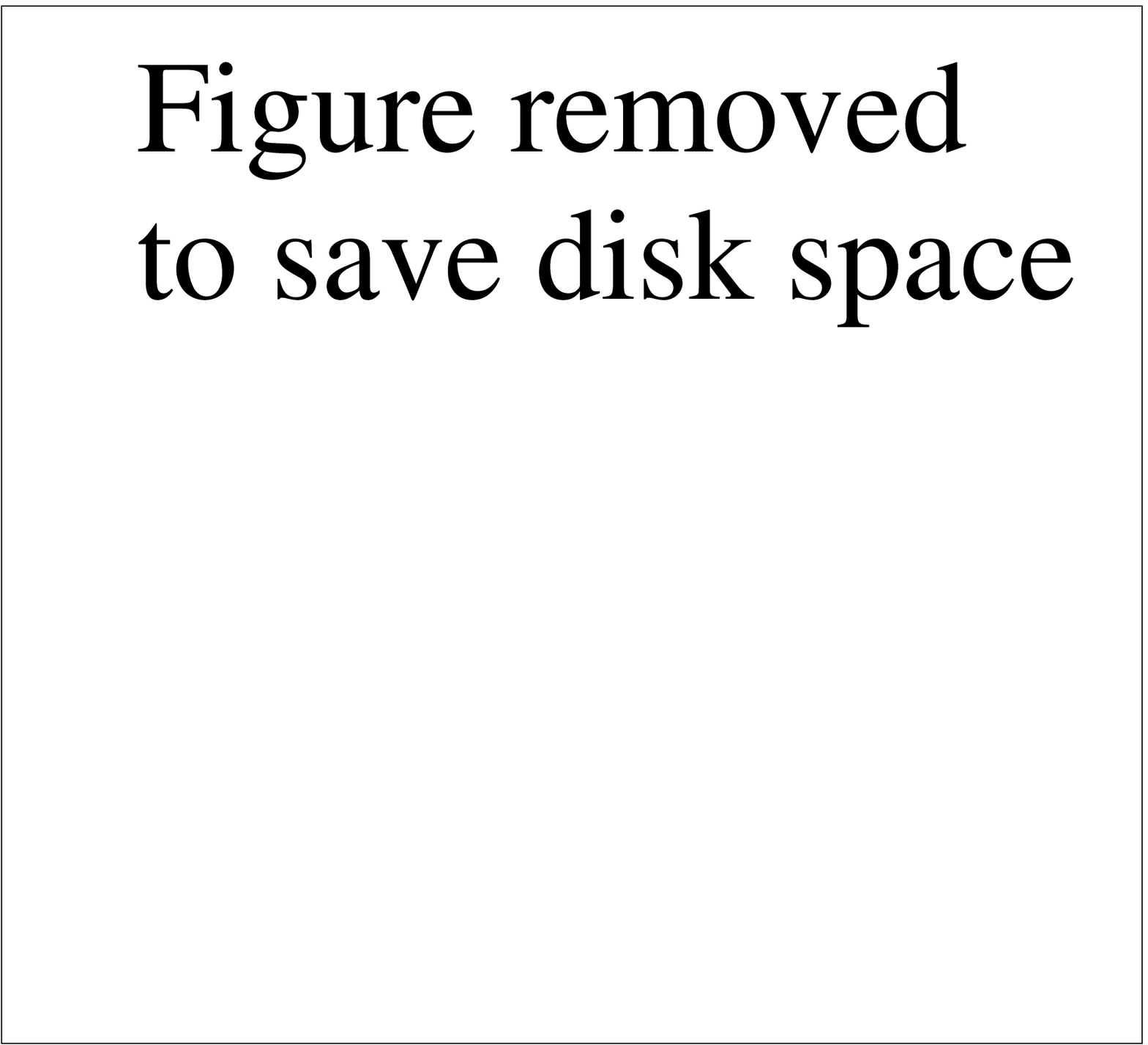}{0.0in}{90.}{200.}{320.}{50.}{0.}
\caption{Top: Histogram of \nh\ derived through spectral fits of the 209 brightest cluster members. While the distribution is peaked at 2.8$\times 
10^{22}$cm$^{-2}$ at least two cluster members have an order of magnitude more absorption. Bottom: Histogram of \nh\ for the 157 cluster members with $<$ 30 counts.
\label{nh}}
\end{figure}

\begin{figure}[t]
\plotfiddle{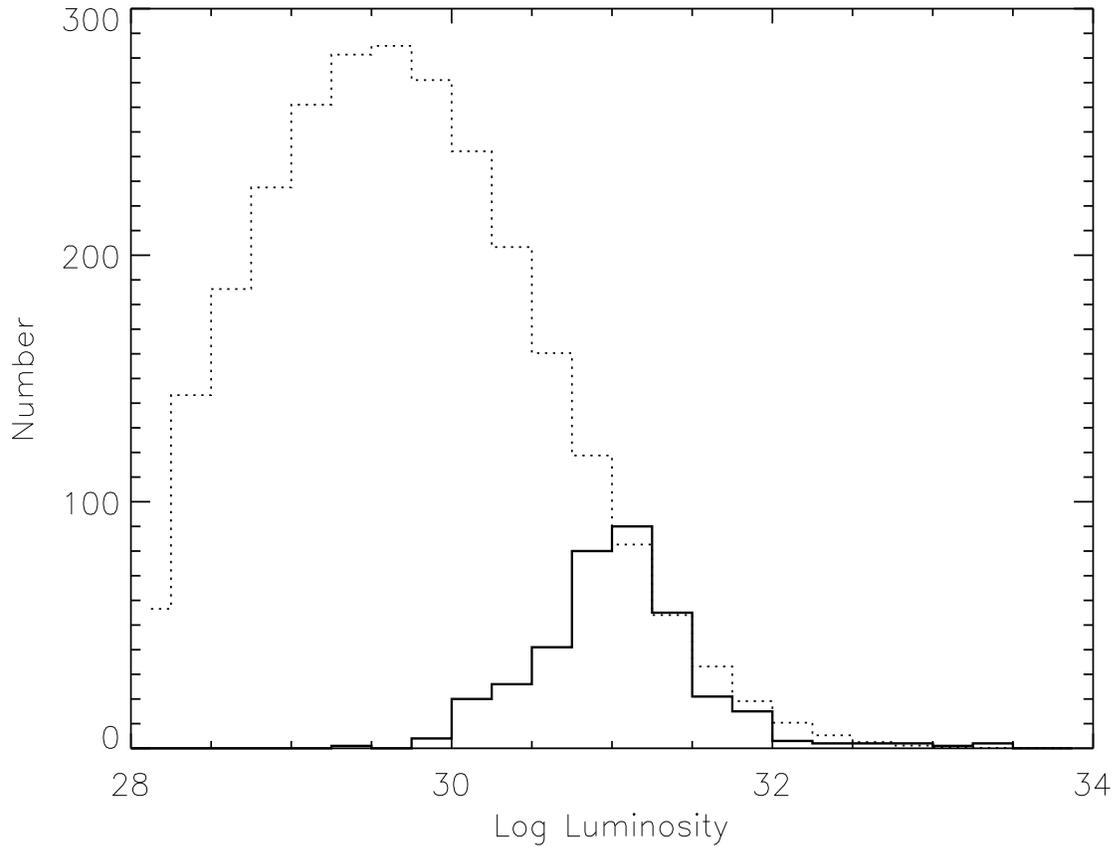}{0.0in}{90.}{400.}{520.}{-50.}{0.}
\caption{Histogram showing the X-ray luminosity distribution in RCW
38.  The solid line show the distribution of 365 candidate cluster members with spectral fits. The dotted line is the expected full distribution of X-ray luminosities assuming a log-normal distribution (After Feigelson \e 2005).
\label{xlf}} 
\end{figure}

We summarize all the spectral data in Figures~\ref{kt}, \ref{nh} and \ref{xlf} which show the histograms of the derived plasma temperature, source extinction and X-ray luminosity function (XLF) respectively. 
Focusing first on the cluster member sources with 30 counts or greater, the typical plasma temperature is similar to that of stars in the ONC (Getman \e 2005). We derive a mean (rejecting
outliers\footnote{Since the mean temperature is about 2 or 3 keV and the total fit range is 0-50 keV the outlying values are vary non--symmetric.  A high outlier can have a value more than 45 keV above the mean, while a low outlier is within 3 keV of the mean.  Hence, we reject 5 $\sigma$ outliers (high and low) before calculating the final mean.} plasma temperature of 2.9 keV (MAD= 0.15).
The bulk of the sources are between 1 and 4 keV with a tail off to higher temperatures. The \nh\ column is symmetrically distributed among the sources with an outlier resistant mean of $2.8\times 10^{22}$ cm$^{-2}$ and a MAD of 
0.10. If we include the fainter members, we find that these tend to be more absorbed with \nh=$4.3\times 10^{22}$ cm$ ^{-2}$ and a little cooler with $<$kT$> \sim 2.0$~keV.  

In Figure~\ref{xlf} we plot both the observed
XLF and the nominal ``complete'' XLF of the ONC, which was parameterized by Feigelson \e (2005) as a log-normal distribution with $<$log L$_x>  \sim 29.3$ and $\sigma \pm 1.0$.\footnote{since we are using photon energies from 0.3keV to 8.0 keV we are using the formulation for total luminosity from Feigelson \e (2005).}  We normalized the log--normal distribution to fit the RCW~38 bins with log L$_x > 31.0$.  The RCW~38 XLF becomes incomplete around log L$_x \sim 30.75$ and is essentially cutoff near log L$_x \sim 30.0$. 
For the case of RCW~38 at 1.7 kpc with a mean log \nh=22.4 the estimated point source sensitivity limit is log L$_x \sim$ 30.1 consistent with the empirical result from Feigelson \e (2005).

\subsection{Diffuse Emission} 
Wolk \e (2002) reported the detection of extensive and bright diffuse emission throughout RCW 38.  In light of the estimate of over 1000 undetected cluster members, it is worthwhile to briefly revisit here whether the diffuse emission may be simply those undetected sources. We find this explanation for the diffuse emission unlikely for several reasons. First, one can attempt to count up the missing photons in Figure~\ref{xlf} by assigning the mean number of missing counts in the first incomplete bin to be 4 (This number is chosen, since sources as faint as 5 counts on axis are generally detected.  It is clearly an over estimate since the mean value will be lower than the maximum) and then summing the counts from each bin.  The result is about 1000 counts, which is significantly less than 6200 counts detected in the diffuse emission.
Second, the region of emission extends about 2\arcmin\ ($\sim$ 1~pc) in the
southeast--northwest direction and about 3\arcmin\ ($\sim$ 1.5~pc) in the northeast--southwest direction (see Figure~2 of Wolk \e 2002). 
Ninety percent of the cluster members are within 1\arcmin\ of the cluster center and there are essentially no members outside of 2\arcmin.  Third, the spectrum of the diffuse emission outside of the cluster center is very hard with temperatures greater than 10 keV if fitted as a thermal spectrum and more accurately fit as a power-law.  This is not the result of the sum of thermal spectra.  It is likely that flux from unresolved point sources in the core contributes significantly to the emission detected there and may account for the thermal nature of the diffuse emission measured in the core.  But stars cannot account for the diffuse emission more than 0.15~pc ($\sim$15\arcsec) from the cluster center.

\section{Infrared Properties of X-ray Sources} 

\begin{figure}[t]
\plotone{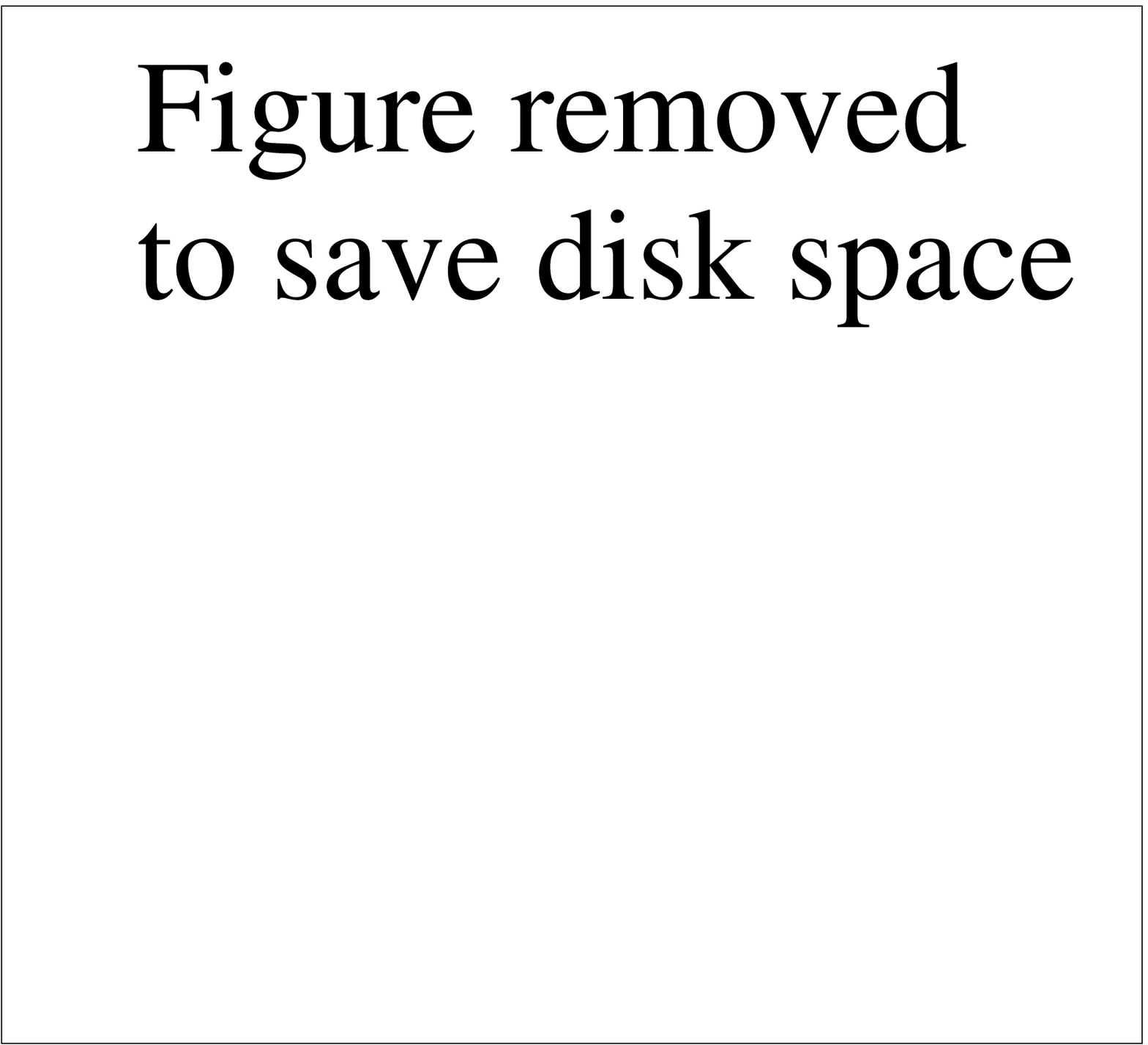}
\caption{Composite J (blue), H (green) and K$_s$ (red) band VLT near-infrared image of RCW~38.   The overexposed bright star is the O5 star IRS~2, and the overexposed ``X'' shaped nebulosity at $\sim$ 08h59m04s --47d30m40s near the center of the image is the IRS~1 ridge (Smith et~al.\ 1999).  The grayscale version of this figure is only the K$_s$ band image. 
\label{VLT_COLOR}} 
\end{figure}

Since the RCW~38 region is optically obscured there is little optical data on the X-ray sources.  The region has been explored in the near-IR however.  A search of the 2MASS point source catalog returns about 2500 sources in the X-ray field.  About 500 of these have photometric errors of $< 5\%$ in all 3 bands.  
However there are only 16 such 2MASS sources in the central 2\farcm5 $\times$ 
2\farcm5. This is because most sources are embedded in extended emission and cannot be properly measured by 2MASS with its 2\arcsec\ nominal resolution.
Because of this, the central region of RCW~38 was observed by Very Large 
Telescope (VLT) in November 1998 as part of the first light observations on VLT 
- UT1 (ESO Messenger 1998, 94, 7).
Data in J, H and K$_s$ bands were obtained using the ISAAC imager with exposure times of 320, 320 and 210 seconds, respectively (Figure~\ref{VLT_COLOR}).
The field measures 2\farcm5 $\times$ 2\farcm5 and was reduced to colors and magnitudes using the standard apphot  and photcal packages in IRAF\footnote{IRAF is the Image Reduction and Analysis Facility, a general purpose software system for the reduction and analysis of astronomical data. IRAF is written and supported by the IRAF programming group at the National Optical Astronomy Observatories (NOAO) in 
Tucson, Arizona. NOAO is operated by the Association of Universities for 
Research in Astronomy (AURA), Inc. under cooperative agreement with the National Science Foundation.}.   There were 447 sources detected in all three filter bands. Minimum photometric errors reach 5\% at J, H and K$_s$ magnitudes of 17.6, 18 and 17.1 respectively.  The errors were dominated by variability in the nebular background.  The 5 $\sigma$ level detections had measured magnitudes of 
19.8, 19.0 and 18.0 in J, H and K$_s$ filters respectively.  
 
There were 36 sources detected in one or two of the VLT filters, but not all three.  This usually occurred when the source was highly obscured and no H and/or J band detection was made. If there was an X-ray source coincident with one of these objects, we estimated magnitudes in the filters in which they were detected by calculating the mean conversion from instrumental magnitude to the standard scale for the 447 sources which had good IR colors and applying this linear factor to the observed instrumental magnitude.
The resulting magnitude estimates have errors of about 20\% since no color term could be applied.  We performed astrometry on this image by matching against 10 X-ray sources spread around the field using the WCStools program imwcs (Mink 2001) which automatically finds stars in an image, matches them to stars in a reference catalog and computes the relation between sky coordinates and image coordinates.  Typical offsets were $<$ 0\farcs3. 
Direct comparison of 2MASS and VLT positions for sources near the edge of the 
VLT field showed similar offsets although 1\arcsec\ offsets were seen in some cases.  We then matched all the X-ray sources in the central region to this VLT catalog. 
For VLT sources, our matching procedure was to match to the nearest source, rejecting all matches with offsets $<$ 1\arcsec. 
The median offset is 0\farcs45.
Outside the central pointing we matched sources against the 2MASS catalog.
The maximum offset between X-ray and 2MASS IR positions allowed in matching was 
1\farcs5. This added error budget was due to the fact that 2MASS was used primarily for sources away from the $Chandra$ aimpoint, hence there is larger positional uncertainty in these $Chandra$ sources.  2MASS colors and magnitudes have typical errors of less than 20\% at magnitudes of about 17, 16 and 15 respectively with errors of about 5\% for sources which are one magnitude brighter than these limits.

Near IR matches were found for 349 of the X-ray sources listed in
Table~\ref{Sources}.  Of the 360 cluster members, 294 have near--IR matches, these are listed in Table~\ref{tbl-irmembers}.  Column 1 of this table is the X-ray source number.  Column 2 is the 2MASS name or a VLT name (J2000) depending on which data are used. Column 3 is the offset between the X-ray and IR positions. Columns 4-9 are the J, H and K$_s$ band observations and errors. Column 10 indicates the A$_V$ as calculated later in this section. 
Column 11 indicates the type of emitting surface assumed in the calculation of 
A$_V$ -- this is either a cool M star photosphere (M star), a higher mass star photosphere (bright) or a star + disk combination (Bright).  Column 12 contains the 2MASS flags for the 2MASS sources.  For the VLT sources, the notation faint is used to indicate sources which were not detected in all three bands and hence subject to large errors due to the lack of color information.   Table~\ref{tbl-irnon} lists the same information for 55 X-ray sources which do not appear to be cluster members. We determine extinctions to 36 of these; none have near IR-excesses which require a disk.

\begin{figure}[h]
\plotfiddle{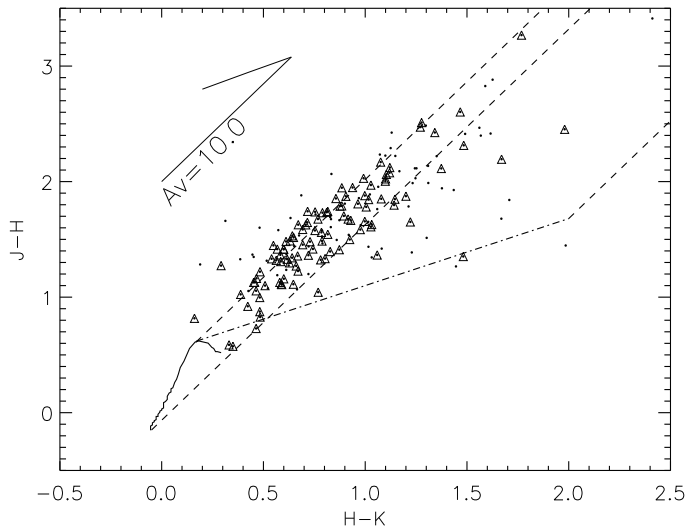}{0.0in}{90.}{250.}{330.}{50.}{50.}
\plotfiddle{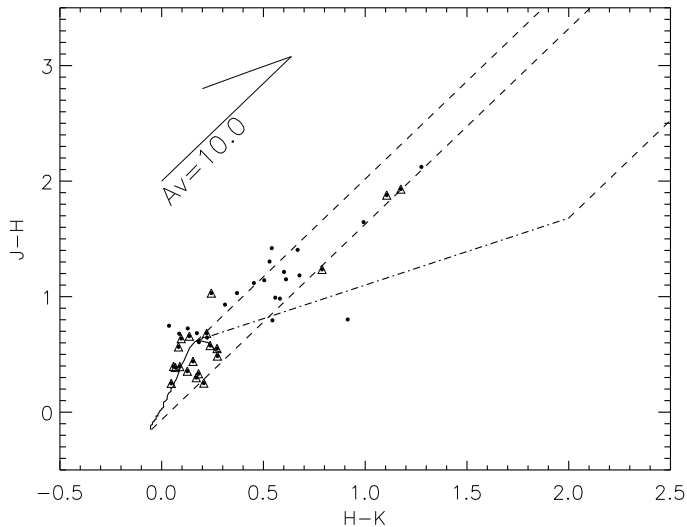}{0.0in}{90.}{250.}{330.}{50.}{50.}
\caption{Infrared color-color diagrams (J-H vs. H-K) for counterparts to all 
X--ray sources in the RCW~38 $Chandra$ field.  The solid line indicates the main sequence, the dashed line running parallel to the reddening vector and the dot-dashed line is the cTTs locus (after Lada \& Adams 1990 and Meyer \e 1997). A reddening vector indicative of A$_V = 10$ is marked.  Sources marked as points have errors $<$ 0.2 magnitudes.  Sources marked also with triangles have errors of $<$ 0.05 magnitudes at K$_s$. Top: sources which are candidate members based on their X-ray quartiles.  Bottom: X-rays sources which are not candidate members.}
\label{jh-hk_x}
\end{figure}

There are 55 matches to the 81 non-members as determined in \S 3.
The color-color diagrams for the cluster members and non-members are shown in
Figure~\ref{jh-hk_x}. In these plots, we restrict the sources to those with errors in J and K$_s$ of $<$ 0.2 magnitudes and highlight sources with errors of less than 0.05 magnitudes. 
Most non--member sources appear to be unobscured blackbodies, although two are obscured by A$_V \sim 10$.   Among the cluster members, none appear to be normal unobscured photospheres.  Most of the X-ray sources which are candidate cluster members appear quite obscured. Twenty-six of the 184 (14\%) IR sources with errors of $<$ 20\% lie within the classical T Tauri star (cTTs) region of the color--color diagram as described by Meyer \e (1997).  
If we restrict ourselves to the 104 X-ray members with $<5\%$ errors, 
9 stars lie in the disk region. While H$-$K$_s$ is a not very sensitive to disk excess (Lada \& Lada 2003), this is still a very small disk fraction for X-ray sources in such a young cluster.

If the IR data are of high enough quality ($err_{mag} < 0.2$) they allow a tentative extinction estimate.  We simply calculate the amount of reddening required to move the object from zero reddening to the observed location.  The zero reddening location is different for a star with and without a disk and varies depending on spectral type for pure photospheres. 
To calculate the extinction we first assume the least obscured case, that all the stars have disks and deredden the photometric colors until they intersect the cTTs locus.  This is the only possible solution for stars outside the pure reddening box (Lada \& Adams 1992).  For stars inside the box two disk solutions exist where the extinction correction crosses the main sequence. 
One of these solutions roughly corresponds to the star being an M star the other with the photosphere being warmer. We measure the extinction for stars inside the box for each hypothesis (disk, M star, higher mass (HM) star). Thus, we determine up to three possible extinctions for each star. For stars which lie either below the cTTs locus or to the left of the main--sequence, no estimate of extinction is possible. In addition there are stars outside the reddening box due to observational errors, not disks.  If the dereddened star is within 
2$\sigma$ of the reddening box on the low mass side of the main sequence (high
H$-$K) we calculate the M star reddening.  We do not calculate the reddening of the HM star hypothesis because the angle between the
ZAMS and the extinction curve is very shallow leading to large errors.
If the dereddened star is outside the reddening box on the high mass side, falling near the elbow of the main sequence in the color--color diagram then no additional reddening is possible beyond the disk/M star hypothesis which yield the same result.  We label these ``GK stars'' in the table since this is their inferred spectral type on the main sequence.

The IR-extinction of stars which are bright X—ray sources can be cross-checked against the \nh\ column derived from spectral fits to the X-ray data.
To start this process we use the relation of \nh\ to A$_V$ recently derived for $Chandra$ ACIS-I data by Vuong (2003) of \nh = A$_V \times 1.6 \times 10^{21} {\rm cm}^{-2}$ and find which model hypothesis (disk, M star or HM star) minimizes the difference in the A$_V$ estimates. The hypothesis which most closely matched the measured \nh\ absorption column was chosen as the probable source extinction in
Table~\ref{tbl-irmembers}. 
In about half the cases the difference between the IR determined A$_V$ and the 
A$_V$ inferred from \nh\ was $<$1 magnitude. However, in about 20\% of cases the difference was greater than 5 magnitudes.  This large scatter led to some concern about the appropriateness of the \nh\ to A$_V$ relation that was used.
The mean difference between A$_V$ calculated using the infrared extinction correction method and A$_V$ calculated using the Vuong relation is 2.62 magnitudes, (MAD=0.35).  

\begin{figure}[h]
\plotfiddle{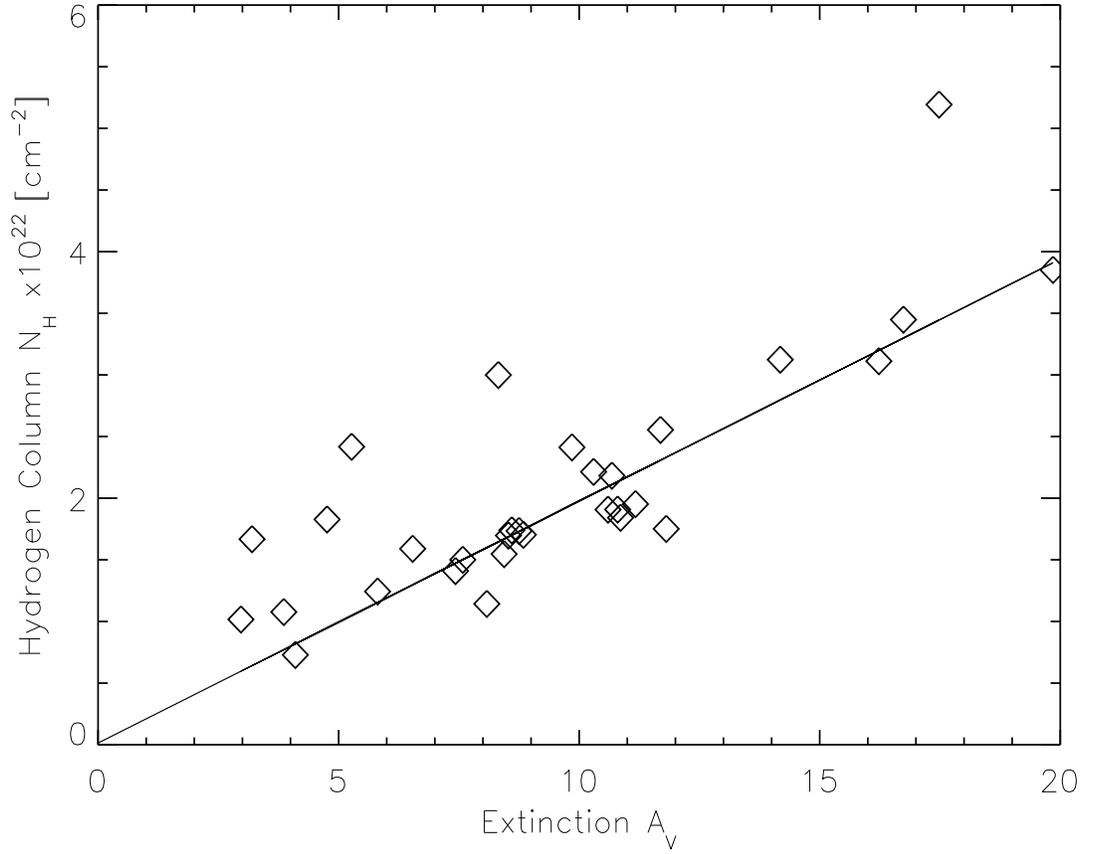}{0.0in}{90.}{400.}{520.}{0.}{-1.0in}
\caption{Plot of the hydrogen column as derived by X-ray spectral fits versus derived near-IR extinction.  The line is the best fit line passing through the origin \nh\ = A$_V$ $\times 2.0 \times 10^{21}$ cm$^{-2}$.}
\label{nh_av}
\end{figure}

To assess and refine the formulation of \nh\ to A$_V$ relation we fitted the extinction, as determined above, to the measured \nh\ values.  We restricted the fit to 31 bright X-ray and IR objects with a magnitude error of less than 5\% in J and an uncertainty of less than 30\% in \nh.  
Using a regression fit which rejected outliers we find a relation of \nh\ = 
A$_V$ $\times 1.7 \times 10^{21}$ cm$^{-2}$.  Forcing zero extinction at zero column of \nh\ we get a best fit of \nh\ = A$_V$ $\times 2.0 \times 10^{21}$ 
cm$^{-2}$ that is shown in Figure~\ref{nh_av}.  These are intermediate to results derived using $ROSAT$ data (Ryter 1996) and those of Vuong \e
(2003). Using \nh\ = A$_V$ $\times 2.0 \times 10^{21}$ cm$^{-2}$ the mean difference between A$_V$ calculated using the infrared extinction correction method and A$_V$ calculated using the \nh\ relation drops to 0.28 magnitudes.  The mean absolute deviation of this fit is 0.29, consistent with no offset.

At this point we revisited the choices we made amongst the 3extinction possibilities of each star to verify that the selection made when we assumed \nh\ = A$_V$ $\times 1.6 \times 10^{21}$ cm$^{-2}$ was still the best fit to the updated \nh\ to  A$_V$ relation. We found that in all cases, the extinction as derived using the infrared method was insensitive to the choice of \nh\ to A$_V$ conversion despite the fact that the conversion is used for vetting the models.  This was because the differences in the possible near-IR extinction hypothesis are more significant than the difference modest 20\% difference between the \nh\ to A$_V$ relations.

The extinction to 160 X-ray selected cluster members was determined via infrared photometry.  Sixteen of these stars have
K-band excesses indicative of disks based on their infrared colors.  
Most of the remainder had multiple viable extinction solutions.
Using the X-ray absorption as a guide, 
33 of these are most consistent with disks for a total of 49 (31\%).    
Thirty-five appear to be obscured M stars without K-band disk signatures while 
49 appear near the ``elbow'' of the main sequence and while no disk is in evidence there is little which can be determined about the photosphere either.
The remaining 28 stars have colors too blue to be K or M stars or have optically thick disks at K$_s$ band.

\begin{figure}[h]
\plotfiddle{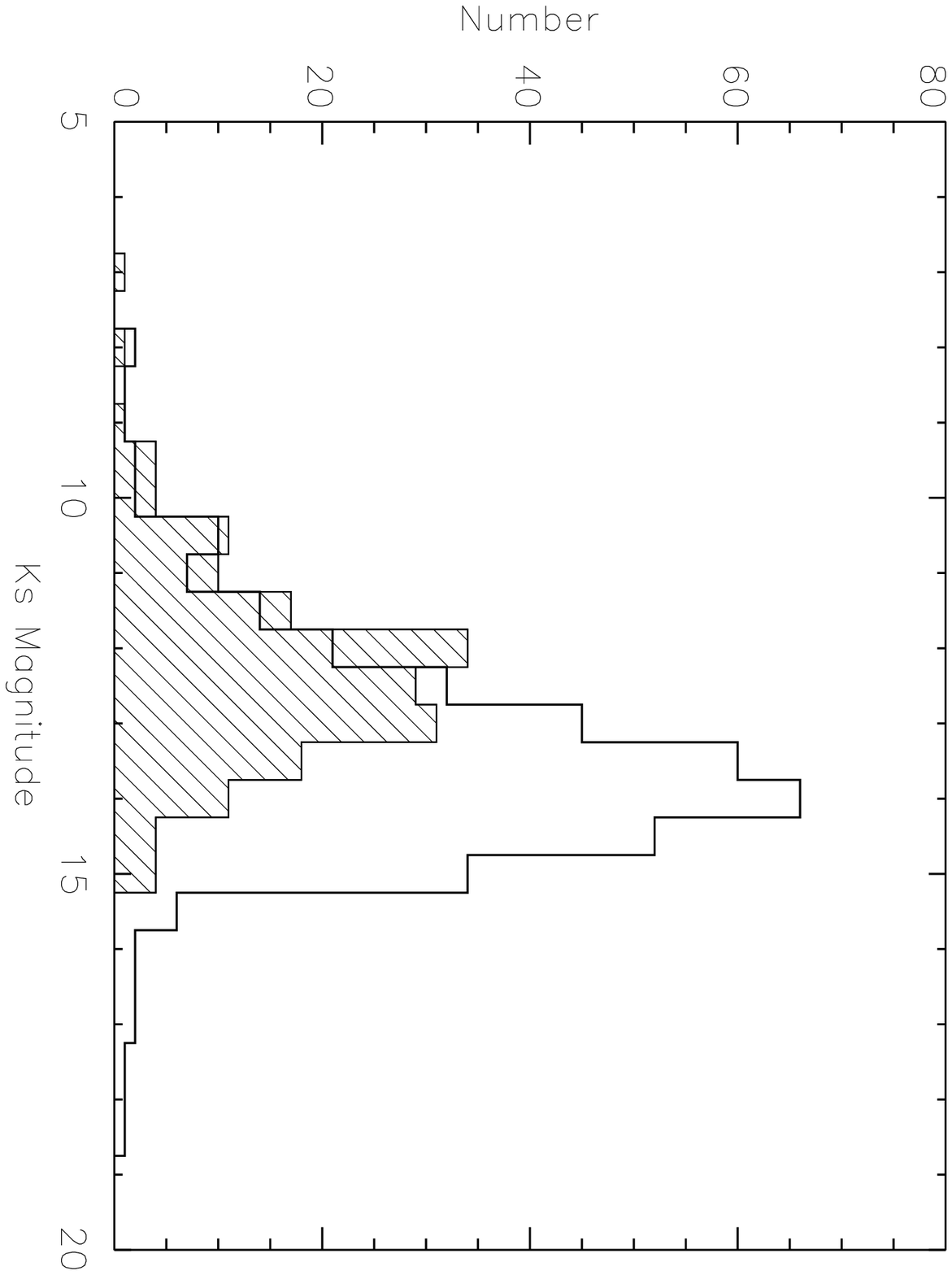}{0.0in}{90.}{270.}{370.}{00.}{-1.0in}
\plotfiddle{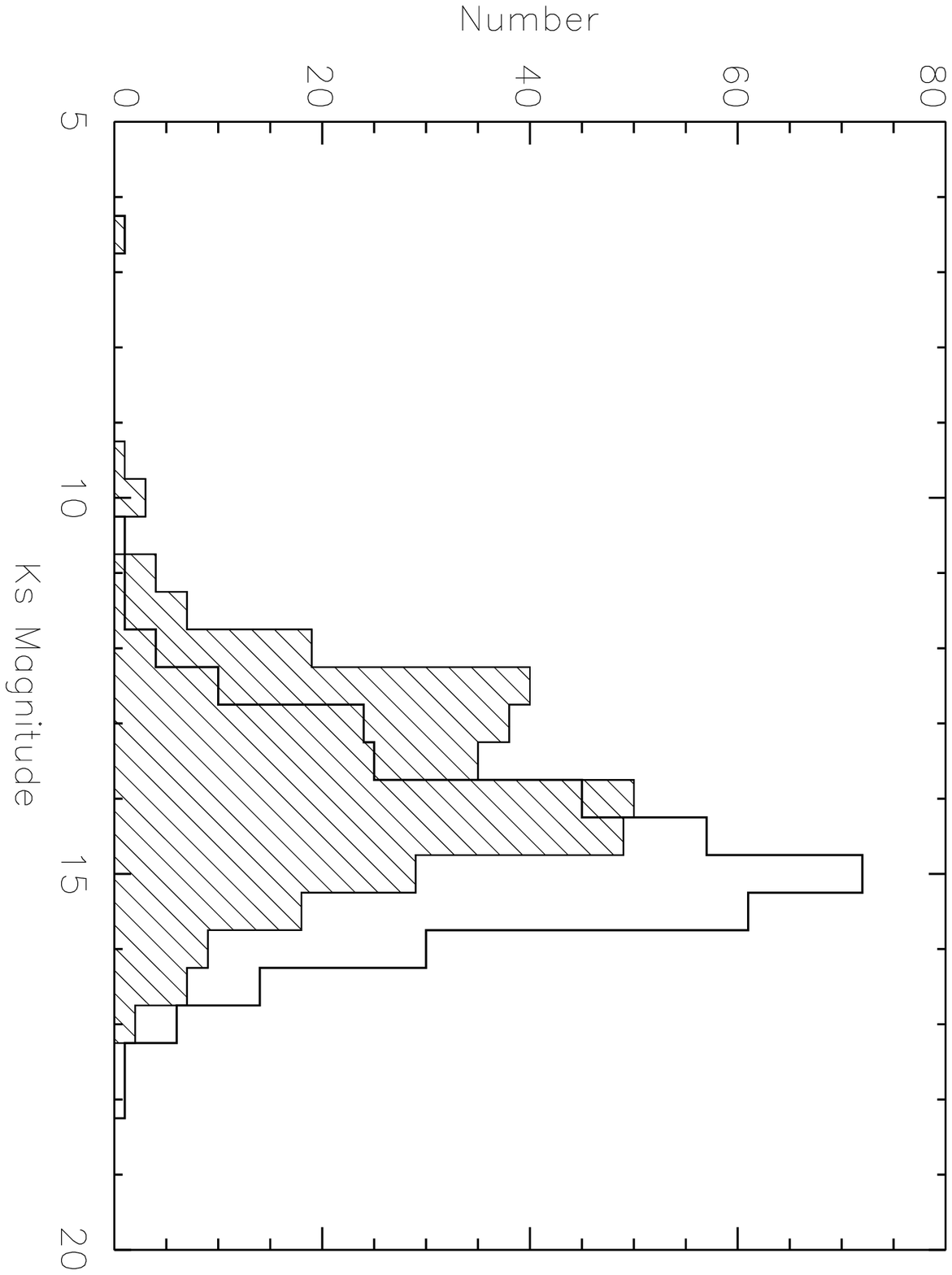}{0.0in}{90.}{270.}{370.}{00.}{0.}
\caption{Plot of the KLF of the X-ray detected members of RCW~38.
Open histogram are data as observed.  Hatched histogram includes the extinction correction. Top: X--ray detected cluster members stars. 
Bottom: non--X--ray detected IR sources in the VLT field.}
\label{KLF}
\end{figure}

We create an intrinsic KLF by correcting the observed K$_s$ band magnitude for extinction by A$_k=0.109 \times $A$_V$ (Bessell \& Brett 1988).  
In Figure~\ref{KLF}, we find the KLF is peaked at K$_s$ = 12.  At a distance modulus of 11.15 this is an absolute K magnitude of about 0.85. 
Masses are estimated using the theoretical isochrones of Siess \e (SDF; 2000) with metallicity =0.02 and no convective overshooting and a cluster age of 0.5
Myr.\footnote{All manner of evolutionary tracks are estimates and concerns about accuracy of the estimates become larger at younger ages (see Hillenbrand \& White 2004 for a full discussion).  Initially, the age is estimated as 0.5 Myr based on the degree of embeddedness of the sources, but we expect a range of ages as seen in the ONC.   We use the Siess \e tracks because of their large mass coverage, especially at high masses.} Metallicity was varied from 0.03 to 0.01 with only a 3\% effect on the K$_s$ band magnitude at 2.7 M$_\odot$.  The error induced by the age estimates, which range between 0.5 and 1.0 Myr was almost 50\%. At an assumed age of 0.5 Myr a star with absolute K$_s$ magnitude of 0.85 has a mass range of about 1.05 M$_\odot$. 
Completeness appears to become an issue at K$_s \approx 13$ which corresponds to 0.5 M$_\odot$.  The ratio of non-X-ray stars per magnitude with corrected K magnitudes between 10 and 12 and those with corrected K magnitudes between 13 and 14 is about 1:4.  The same ratio for the X-ray detected sources is about 1:2. Thus, we detect about 50\% of the stars at an extinction corrected K$_s \approx 13$ (0.5 M$_\odot$).  This is consistent with the prediction given in \S 3.  We have 50 detections at 14$^{th}$ magnitude in K$_s$.  
These sources have masses ranging from 0.25 M$_{\odot}$ down to the brown dwarf limit. 

\subsection{Other IR sources in the VLT field.}

Examination of the VLT non--X--ray detected sources can be used to analyze the completeness of the X--ray data.  Since the VLT image is centered near the core of a very dense cluster with a dense molecular cloud behind it we expect the stellar sources in the VLT field to be dominated by the cluster members.
Using the galactic models of Wainscoat \e (1992) we expect minimal contamination, perhaps 6 sources brighter than K$_s$ = 17 in this small field of view. None of the 129 X-ray sources in the VLT field of view has been excluded as a cluster member based on its X-ray properties.

\begin{figure}[h]
\plotfiddle{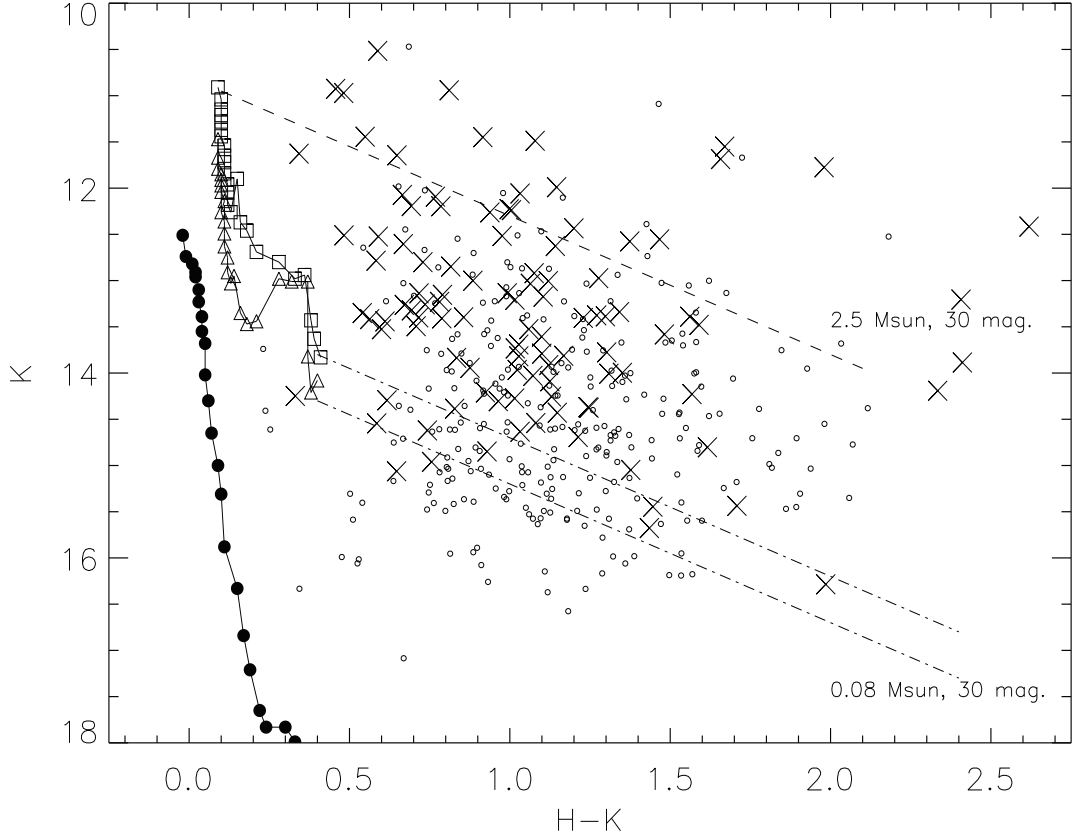}{0.0in}{90.}{400.}{520.}{-50.}{0.}
\caption{
Infrared color-magnitude diagram for all sources in the VLT field.  X-ray sources are indicated as ``X's'', non--X--ray sources are indicated with small circles. The 0.5 Myr (open squares) 1.0 Myr (open triangles) and ZAMS (filled circles) isochrones for 2.5 M$_\odot$ to 0.08 M$_\odot$ are plotted (Siess \e 2000).  
Extinction of 30 visual magnitudes for a 2.5 M$_\odot$ star is indicated by the dashed lines. The dot dashed lines indicate 30 visual magnitudes for 0.08 M$_\odot$ stars at ages of 0.5 Myr (upper) and 1.0 Myr (lower).}
\label{CMD}
\end{figure}

There are a total of 482 stars in the VLT field with good colors in J, H and K$_S$. Of these, 129 are X-ray sources.  
We assume the bulk of the remaining 353 are cluster members. 
We note that about 25\% of the IR sources were detected in the X-ray observation (after a small correction for VLT sources detected in less than 3 bands).  This is somewhat at odds with the prediction of a 16\% detection rate based on ONC data (Feigelson \e 2005).  The VLT data reach K$_s$ of 17.1.  This should be complete to 0.1 M$_\odot$ even through 30 magnitudes of visual extinction according to the SDF tracks at 0.5 Myr and 1.0 Myr. However, very low mass stars may be 0.5 magnitudes fainter than this and 20 Jupiter mass brown dwarfs similar to those detected in the ONC may be 3 magnitudes fainter than the faintest sources of stellar mass (Hillenbrand \& Carpenter 2000). Thus the VLT image is somewhat incomplete. The IR color--magnitude shown in Figure~\ref{CMD} uses SDF isochrones set at 0.5 Myr, 1.0 Myr and the ZAMS. The dashed lines indicate 30 magnitudes of visual extinction. X-ray sources are noted with ``X'' and non-X-ray sources with open circles. We find about 10 of the 482 stars are too faint and unobscured to be cluster members. Thus, there is some contamination by non-cluster members in the VLT dataset.  If we take the 16\% detection fraction as a minimum fraction of cluster members detected in X--rays and 25\% as the maximum, then the total cluster membership is between 1450 and 2400. 

\begin{figure}[h]
\plotfiddle{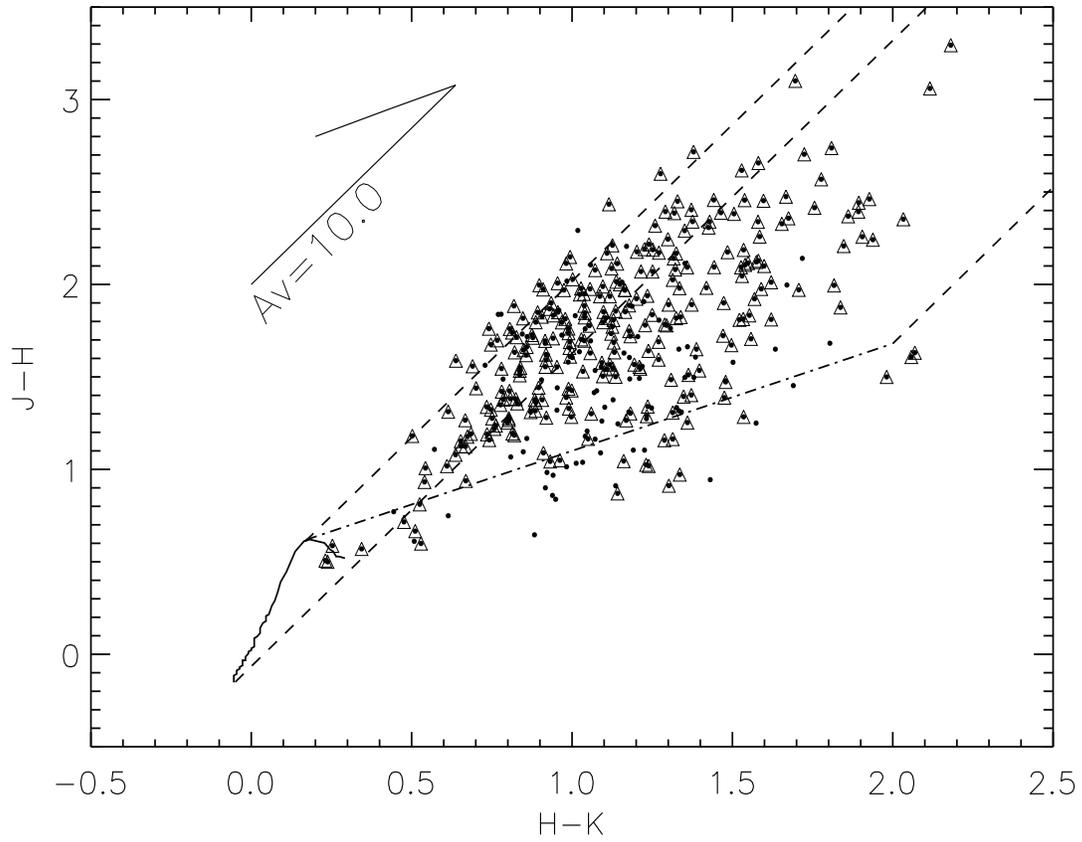}{0.0in}{90.}{400.}{520.}{-50.}{0.}
\caption{Infrared color-color diagram (J-H vs. H-K) for VLT sources without 
X--ray counterparts in the RCW~38 {\it Chandra} field.  Symbols are the same as 
Figure~\ref{jh-hk_x}.}
\label{irccd_nox}
\end{figure}

In Figure~\ref{CMD}, it appears the X-ray sample is not uniform compared with the non-X-ray sample.  The X-ray sample seems more sensitive to very embedded, high--mass sources and less sensitive to embedded lower mass sources. 
The 353 VLT sources not detected in X-rays are indicated in the color--color diagram shown in Figure~\ref{irccd_nox}. The minimum extinction to each star was derived using the same procedure as in the previous section.  
One--hundred and seventy--three stars (49\%) in the non--X--ray group needed to be modeled as stars with disks.   This fraction is much higher than the 14\% disk fraction among the X-ray sources.  
Further the 49\% disk fraction is a lower limit.  The remainder of the stars may or may not have disks as the near-IR is only sensitive to the hot part of the inner disks. Without additional data (e.g. X-ray spectra or
mid-IR photometry) it cannot be determined which of the other 51\% have disks.

This indicates that the X-ray sample is biased away from stars with inner disks.  
There are two possible, non--exclusive, explanations for this.  First, the disk itself will absorb X-rays.
Second, the X-ray production mechanism for stars with disks may be different and lower in luminosity than it is for PMS stars without disks.  These possibilities are discussed further by Preibisch \e (2005). The intrinsic K$_s$ band luminosity is {\em not} the primary difference between the X-ray detected and non--X-ray detected populations. Of the 152 VLT stars detected in X-rays, 8 are fainter than K= 16 and 2 are fainter than K=18.  Out of the remaining 353 VLT detected sources 36 are fainter than K= 16 and 1 is fainter than K=18.  The ratios are statistically indistinguishable. The KLF of the non--X--ray detected sources (bottom of Figure~\ref{KLF}) shows a similar shape to the KLF of the
X-ray detected members.  Finally a two-sided KS test comparing the colors of the brightest 100 X-ray and non-X-ray detected sources found a 4\% probability that they were drawn from the same population.  Even when all sources were included the KS test found a 1\% chance that they were drawn from the same source population.  We conclude that the X-ray sample, while not complete and certainly biased against disks, is not particularly mass biased down to the sub-stellar limit. Only one X-ray source is detected significantly below the nominal brown dwarf limit, Source~281 at 0\farcs68 from VLT085907.79-473122.2, a very likely cluster member since its extinction is about 5.25 A$_V$.

\subsection{Possible O and B stars} 
Using the extinction corrected KLF we performed a survey of the cluster for O and B star candidates. On the ZAMS a star of 2.7 M$_\odot$ is a late B star. On the SDF mass tracks, at 0.5 Myr a 2.7 M$_\odot$ star has an absolute K$_s$ band luminosity of -0.35 as calculated using the on-line SDF model isochrones as discussed in \S6. We chose an age of 0.5 Myr and solar metallicity ($Z$=0.02).
To convert the effective temperatures resulting from the models to colors, we used the conversion table from Kenyon and Hartmann
(1995). Both mass and luminosity were derived from the models using the extinction corrected K$_s$ magnitude of the X-ray selected members as estimated earlier in this section and the distance modulus of 11.15. 

This use of the SDF isochrones leads the identification of 31 X-ray sources as candidate OB stars.  One star (Source 
390) was excluded since it had no measurable extinction and hence is probably 
foreground.\footnote{Source 390 had less than 30 X-ray counts, hence its identification as a cluster member was already on questionable footing.}
Another, Source 435, is also excluded because it is both a very faint X-ray source and somewhat softer than the rest of the group.  Including Source 
221/IRS~2, which has been discussed elsewhere as an $\sim$O5 star (Smith \e 1999, Wolk \e 2002) and Source 251 which is a very absorbed and luminous X-ray source, there are 29 X-ray detected candidate OB stars.

There are 2 additional stars in the VLT field, which while not detected in 
X--rays are still identified as O or B star candidates based on their extinction corrected luminosity.  VLT085905.44-473045.2 (VLT232) is in the wings of several 
X-ray sources which could easily obscure X-ray emission from this source, up to about 30 counts. This would provide a perfectly reasonable log (L$_x$/L$_{bol}$) of -5.89.  It is somewhat surprising that VLT085906.96-473023.2 (VLT453) is not an X-ray source if it is indeed a B star.  While it is in the vicinity of Source 
251, it is 3\arcsec\ away and the background contamination is only about 0.5 counts.  If this is a B star is would be very weak X-ray source, perhaps due to an unusually weak wind. 

\begin{figure}[t]
\plotfiddle{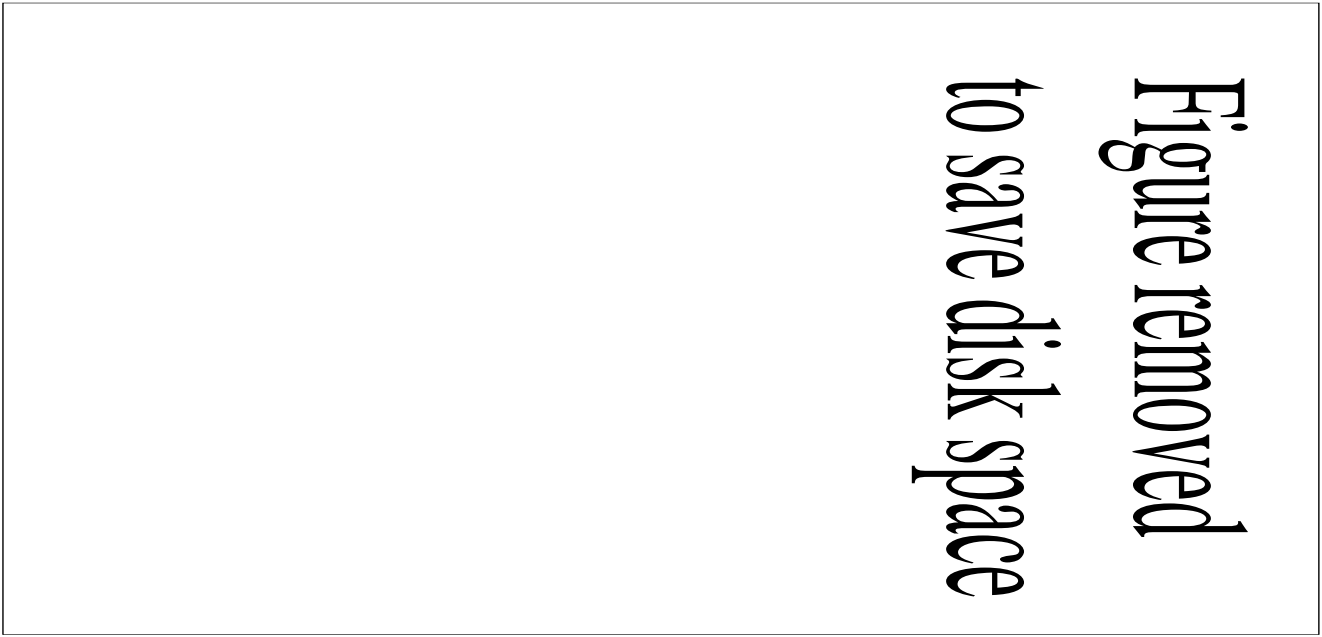}{0.0in}{-90.}{240.}{500.}{-50.}{0.}
\caption{Location of the OB star candidates.  Left: Location of OB star candidates within 8\arcmin\ of the cluster center noted on a 2MASS K-band image. 
The circle indicates 200\arcsec\ from the cluster center.  
The box indicates the VLT field shown in the figure to the right.  
Right:  Location of OB star candidates within 1.25\arcmin\ of the cluster center overlain on the VLT K$_s$ band image. The numbers indicate the X-ray source number, except in the case of the two VLT only sources.  To avoid clutter, only the seconds of declination are listed for the VLT sources.}
\label{OBfig}
\end{figure}

Figure~\ref{OBfig} shows the location of the 31 OB star candidates.
In Table~\ref{tbl-ob} we list the OB stars in RCW~38.
Columns 1--4 of the table list the Source ID, RA, DEC and distance from the cluster center respectively.  Columns 5 and 6 list the magnitudes of visual extinction as derived from the IR 
(col.~5) and X--ray (col.~6) methods. 
The absolute K$_s$ magnitude derived from the measured K$_s$ magnitude, the distance modulus and the infrared extinction is given in column 7.
In general, the infrared extinction is lower than the estimate derived from the \nh\ column.  This occurs for stars fitted with the disk model since we stop the dereddening process when we reach a model consistent with a late type star with a disk.  For the most part the agreement between the extinction measurements (infrared derived and X-ray derived) is very good leaving less than 0.3 magnitudes of variation in the derived absolute K$_s$ magnitude. This contributes about a 25\% uncertainty to the mass estimate. This is usually in the sense of a mass underestimate.  The last 2 columns list the bolometric luminosity (derived from the age and absolute 
K$_s$ magnitude and the relative X-ray luminosity (log (L$_x$/L$_{bol}$)).  The latter values are consistent with known O and B stars. 

The distribution of the OB candidates mimics the roughly symmetrical and peaked distribution of the cluster as a whole. Twenty--one of the candidates are within 111\arcsec\ of the cluster centroid. Four candidates lie more than 300\arcsec\ from the cluster centroid. The most distant candidates tend to have extreme values of  
L$_x$/L$_{bol}$ indicating that they may not be at the distance of RCW 38.
The finding of $\sim$ 31 candidate OB stars is consistent with our estimate for the overall membership cluster as about 1.7 times that of the ONC.
The ONC has about 15 O and B stars (Stelzer \e 2005). Although 31 OB stars is a little on the high side, this number is non disallowed by the radio data (Vigil \e in prep.) since the luminosity is dominated by IRS~2. The large number of candidates may indicate that the 4 off-axis candidates may not be cluster members.  If we assume the cluster is 1~Myr instead of 0.5~Myr the cutoff for O and B stars become more than 0.5 magnitudes fainter (Abs K = 0.32) and would bring the total population of O and B stars in the cluster over 60 which we consider unlikely.

\section{Specific Interesting Sources}
For the analysis presented here, the primary utility of X-ray emission in young stars is as a tracer of youth and to measure the gas and dust along the 
line--of--sight. The former is especially useful, since other, more obvious tracers of youth, such as near-IR emission fade more quickly with age than X-ray emission.
However, several of the X-ray sources, both cluster members and non-cluster members are interesting for their X-ray characteristics.  
In our data set we found two remarkably cool objects, and several sources with very hot plasma and others which were extremely embedded.  We discuss them and the brightest 10\micron\ source below.

\subsection{Very Bright IR sources} 
RCW~38 was identified as a star forming region through the presence of several bright infrared sources. IRS~2 (Frogel \& Persson 1974) is clearly the revealed central O star.  As we discussed previously (Wolk \e 2002), this is coincident with the brightest X-ray source in the field, Source~221. The X-ray count rate is
0.062 counts $\ps$ so the pile up fraction is very small. The X--ray spectrum was not well fitted by any single component thermal model.  Our best fit was to a two--temperature Raymond-Smith plasma dominated by a cool component at 0.77 keV and a 
warm component at 2.5 keV. These temperatures are very consistent with those of 
strong wind sources in the ONC (Stelzer \e 2005). 
Strong wind X-ray sources were modeled by Stelzer \e to have a hydrodynamic wind with many small shocks. IRS~2 has an X-ray luminosity 10\% greater than that of $\theta^1$ Orionis C.  While this star was too bright to be measured in the VLT observations, good 2MASS data are available and demonstrate that IRS~2 is about 5--10 times more luminous than $\theta^1$ Orionis C at K-band.  

IRS~1, the brightest 10 \micron\ source (Frogel \& Persson
1974), appears not to be a point source at all. Rather it is a complex region to the west of IRS~2.  Mid-infrared 10 and 20 \micron\ observations by Smith \e 
(1999) find that IRS~1 is a dust ridge with a number of peaks, including one at the location of IRS~1 and others noted by the letters A--D in Figure~1 of Smith et al.  They propose that the region can be explained in terms of a wind-blown cavity, where the stellar wind from the hot young star 
IRS~2 has cleared a cavity about itself, and the ridge of emission near IRS~1 represents the material which has been swept up into the shell around IRS~2. Vigil \e (in prep.) find that there is a peak in the cm continuum emission within this ridge. They conclude that the density enhancement could lead to future star formation. 

The dust enhancements identified as regions A, B and E by Smith \e are associated with X-ray emission -- sources 187, 197 and 174 respectively.
Source 187 is among the fainter sources with only 18 counts, the absolute K magnitude is very close to that expected of a 0.4--0.5 M$_\odot$ star. 
Source 197 on the other hand is among the brightest 5\% of X-ray sources in the field. Its absolute K magnitude is consistent with that of a 2.2 M$_\odot$ star.  
Both sources appear to be typical thermal plasmas. 
Mid--infrared regions C and D are not coincident with X-ray sources.

Source 174 (08h59m03.67s, -47$^{\rm o}$30\arcmin40\farcs1)\ is the strongest 
X--ray source in the IRS~1 region.  This source is well matched to the westernmost star in a triplet located on the NIR dust ridge that is clearly visible in the center of Figure~\ref{sourcescentral}.  The IR counterpart to Source 174 is highly reddened with a K magnitude near 13, H-K$_s$ $\sim 1.7$ and no J detection.  The visual extinction inferred from the \nh\ column is about 17.  From this we infer a mass of about 2.0 M$_\odot$.  The best fit to the 
X--ray spectrum is a fairly hot 5.1 keV plasma. The high temperature may be indicative of very long (perhaps ten's of stellar radii), low density, magnetic structures in this object supporting the idea that the IRS~1 is in a region of ongoing star formation. Any soft X-ray component would be absorbed so we cannot comment on whether there is a bright soft component to the X-ray emission here.

\subsection {Cool Objects} 
\begin{figure}[h]
\plotfiddle{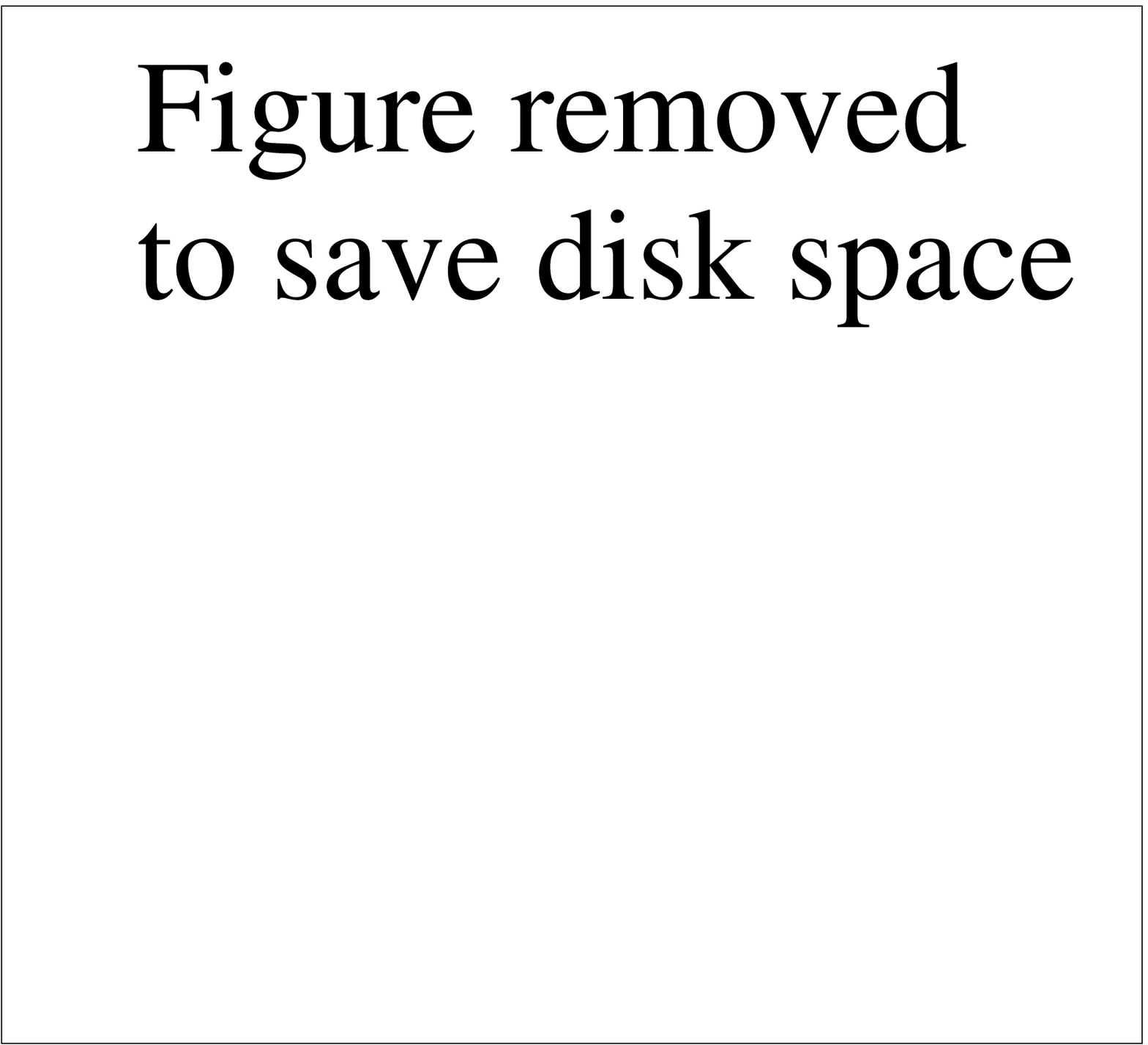}{0.0in}{0.}{270.}{270.}{50.}{-1.0in}
\plotfiddle{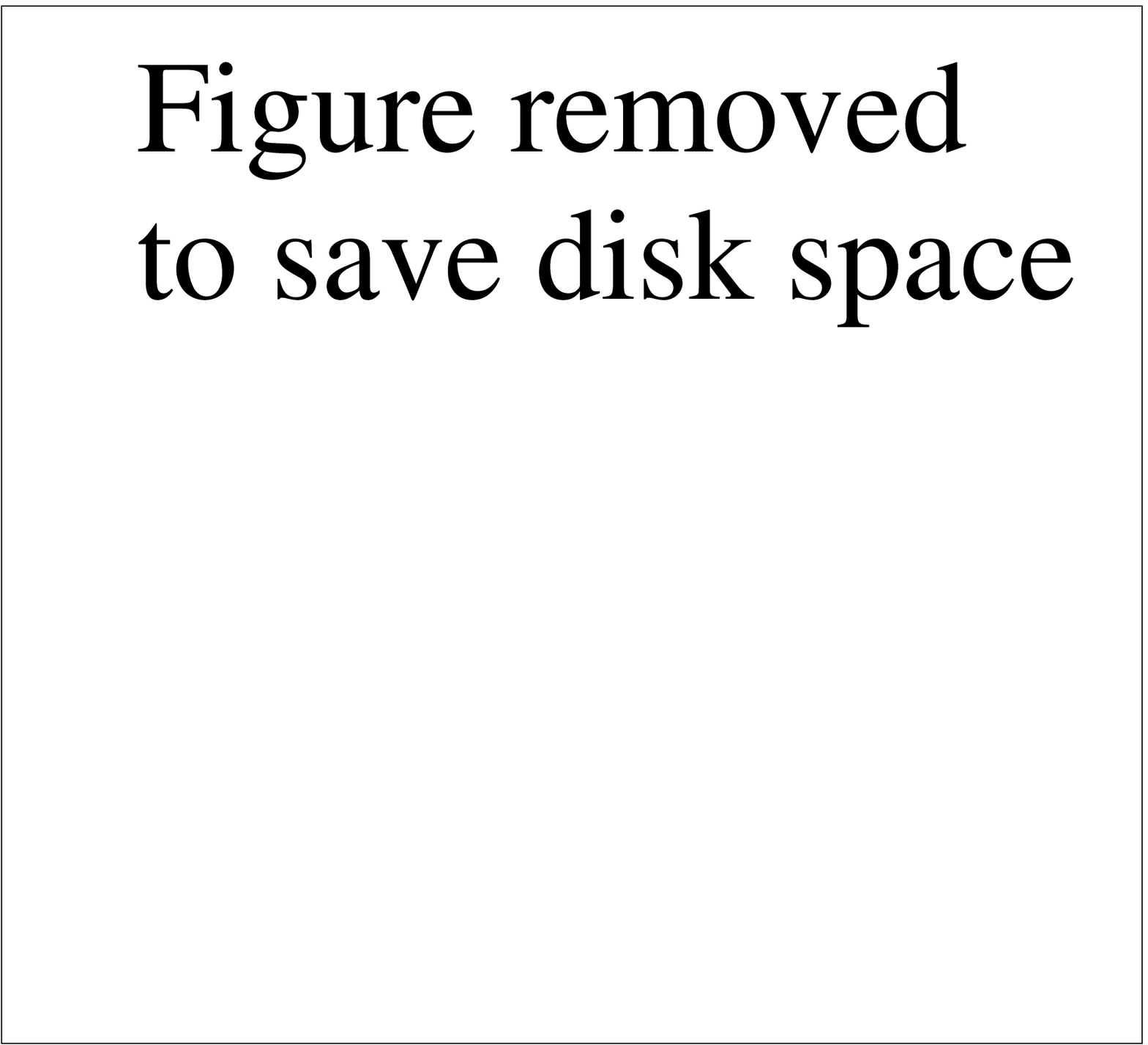}{-1.0in}{0.}{270.}{270.}{50.}{-1.0in}
\caption{X-ray spectra of two cool white dwarf candidates. These spectra are unusual in their lack of flux above 2.5 keV.
They have been fitted with blackbodies with temperatures below 150 eV (histogram), consistent with the emission expected from a degenerate star.}
\label{wd}
\end{figure}

Sources 39 and 447 are both very cool and bright. Both exceed 300 counts with almost all the emission below 2.0 keV. They are the only 2 sources in the data set which are well fitted by blackbodies. Figure~\ref{wd} shows the detailed blackbody fits. Both fits have null probabilities in excess of 0.95. The \nh\ columns fitted to the sources are 6.0 and 7.5 $\times 10^{21}$ cm$^{-2}$ respectively, thus we conclude that they are foreground objects.  The derived blackbody temperatures are 100 and 150 eV, consistent with the emission expected from the surface of a neutron star or, more likely, a white dwarf.  Neither shows evidence of variability.

Both are associated with bright 2MASS sources. 
Source 39 is 0\farcs1 offset from 2MASS085850.39-473319.5 with K=
8.9, J$-$H=0.44 and H$-$K=0.15.  This is consistent with the star lying near the 
G, K, early M stars portion of the color--color diagram with little reddening (less than 0.03 magnitudes at V).  Further this source has Tycho BT and VT magnitudes of 11.93 and 11.40 respectively and a proper motion of 47.1 $\pm2.5$ mas/yr (H{\o}g \e 2000).  The colors are consistent with an unabsorbed K3-4 main sequence star (Koornneef 1983) within 100 pc.  The X-ray source is most likely an unseen degenerate companion.  

Source 447 is 0.35\arcsec\ offset from 2MASS085956.09-473304.3 which has K= 8.1, 
J$-$H=0.33 and H$-$K=0.18, colors that are consistent with a late K type star.  
However the source is also coincident with LS~1223 with V$_{mag}$=11.48, B$-
$V=0.86 and U$-$V=-0.19, consistent with a moderately reddened (E$_{(B-V)}$ 
=0.4) early B star.  This extinction is not consistent with the measured \nh\ column or the IR extinction which are A$_V$ of 3.75 and 3.0 respectively. 
Further, the derived V$-$K$_s$ of 3.4 is consistent with an unobscured K4-K5 star. To obtain consistent colors we need to model source 447 as a K4 star with a young white dwarf companion (which may account for the blue excess) at about 50 pc.

\subsection{Very Hot objects} 
\begin{figure}[h]
\plotfiddle{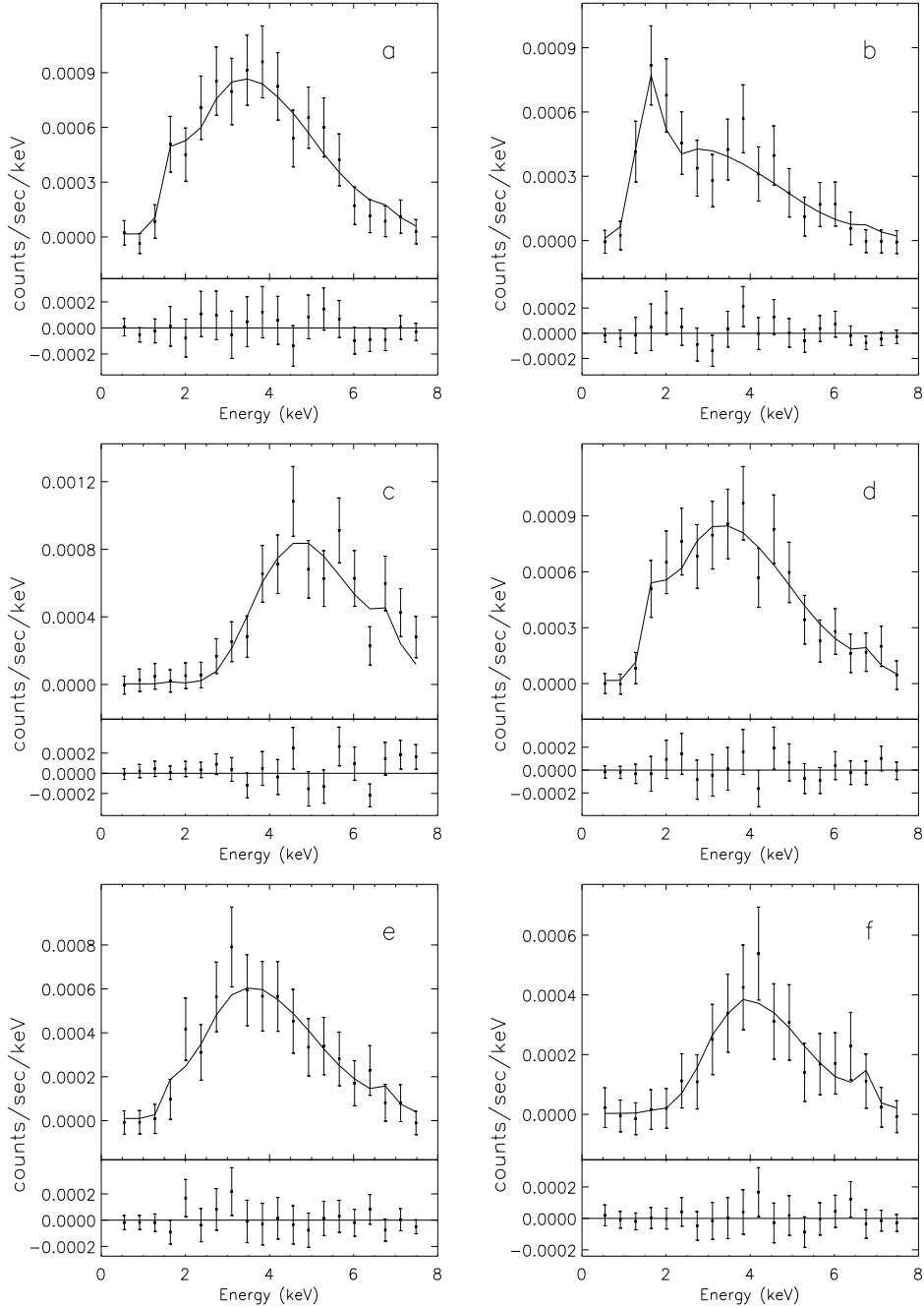}{0.0in}{0.}{370.}{530.}{50.}{-1.0in}
\caption{Spectra of the hottest, most embedded sources in our sample.
Source 56 (a) is fitted with a 14.7 keV thermal plasma with \nh $\sim$ 3.9 
$\times 10^{22}$ cm$^{-2}$. Source 147 (b) is fitted with a 11.0 keV thermal plasma with \nh $\sim$ 2.7 $\times 10^{22}$ cm$^{-2}$. Source 251 (c) is fitted with a 10.7 keV thermal plasma with \nh $\sim$ 2.3 $\times 10^{23}$ cm$^{-2}$. 
Source 149 (d) is fitted with a 9.6 keV thermal plasma with \nh $\sim$ 3.8 
$\times 10^{22}$ cm$^{-2}$. Source 108 (e) is fitted with a 8.9 keV thermal plasma with \nh $\sim$ 5.37 $\times 10^{22}$ cm$^{-2}$.  Source 118 (f) is fitted with a 3.7 keV thermal plasma but a large \nh $> 10^{23}$ cm$^{-2}$. All sources have a reduced $\chi^2$ $<0.6$ except for 251 which has a reduced $\chi^2 \sim 0.95$. Residuals to the fits are shown beneath each fit.}
\label{56147}
\end{figure}

At the other end of the spectrum there are five bright sources which were best fitted by a single temperature plasma with a temperature exceeding 100~MK (kT $> 
8.6$ keV) and in excess of 100 counts so we have good confidence in the fits.\footnote{There are three sources fitted with two--temperature plasma in which the warmer plasma exceeds 100~MK. However, in all three of these cases the emission measure is biased towards the cool component such that there are less than 50 counts in the hot component.}  
Such stars are interesting because the underlying physics required to generate these high temperatures is probably quite different from the physics required to generate the roughly 2.5 keV characteristic plasma seen around most million year old stars.  At a minimum, such temperatures probably require magnetic structures exceeding a stellar radius (Favata \e 2005). Further, the physics required to sustain such plasmas for the entire observing window are probably not related to the loop reconnection seen in flares. We discuss the sources in order of descending temperature.
Two of the five stars flared during the observation, for these stars the high temperatures are at least partially related to the flare episodes. The other three sources showed no evidence of strong flares.
The spectra of all five are shown in Figure~\ref{56147}.
 
{\bf Source 56 --} This source, with about 309 counts, is fitted with the hottest plasma in the field.  
The fit with a 14.7 keV thermal plasma with \nh = 3.9 $\times 10^{22}\persqcm$ is very good with a reduced $\chi^2$ of $< 0.5$. It is not in our VLT field and has neither 2MASS, nor MSX counterpart so it could be an AGN or other exotic object. 
Fits with a power-law yield a slightly poorer fit with reduced
$\chi^2$ of $0.33$ but perfectly reasonable values of \nh =4.05 $\pm
0.07 \times 10^{22}\persqcm$ and $\Gamma =  1.83  \pm 0.21$, indicating it could be powered by synchrotron electrons. However, it is 166\arcsec\ from the cluster center which puts it on the edge of the cluster core.  While it is possible that this is a chance superposition of a background object, its location and count rate make a protostellar hypothesis more likely. 

{\bf Source 147 --} This source, with almost 200 counts is best fitted with an 
11.0 keV thermal plasma with \nh =1.7 $\times 10^{22}\persqcm$ with a reduced  $\chi^2$ of $< 0.6$.  Fits to a power-law yield a poorer fit with reduced $\chi^2$ of $\sim 0.45$ and reasonable values of \nh =0.9$
\times 10^{22}\persqcm$ and $\Gamma =  1.07$ but, the residuals are systematic indicating that the power-law model expects too much flux at the hard end.   Further, since it is only 50\arcsec\ from the cluster center it is within the VLT field of view.  Indeed it is quiet bright in the near-IR with K=11.5 and an A$_V \sim 10$ consistent with the \nh\ measurement.  The absolute magnitude of -0.65 at the 1.7 kpc distance of RCW~38 is consistent with a 3.5 M$_\odot$ star near the birthline (Palla \& Stahler 1993).  The mass of this star could be significantly higher due to the non-monotonic relation between K magnitude and mass in this regime. 

{\bf Source 251 --} This source, with over 280 counts, is the closest to the cluster center among this group, $<$ 22\arcsec\ off--axis.  It is the only one of the five hottest sources coincident with the molecular ring discussed by Vigil \e (in prep.). It lies on the inner edge of the molecular and dust rings.  It is fitted with a 10.7 keV thermal plasma with a very high column, \nh = 2.3 $\times 10^{23}\persqcm$ with a reduced $\chi^2$ of $0.95$. 
Fits to a power-law yield a negative slope or forcing a positive slope yields reduced  $\chi^2$ of $> 9$.  Since the star is highly embedded it is not detected at J band. Our K$_s$ band detection is 14.9 with H-K$_s$ of
1.5. Based on the \nh\ column we estimate A$_K$ to be about 12.5.
This yields an absolute K magnitude of -8.75.  This corresponds to about 25,000 
$L_\odot$ which is similar to that of the Becklin--Neugebauer object (Becklin \& Neugebauer 1967) and consistent with an O star of greater than 10 M$_\odot$.   
This star has a clear flare, about 80\% of the observed flux from the star is associated with that flare.

{\bf Source 149 --} This is a 300 count source 95\arcsec\ off--axis.  
It is outside of the VLT field but coincident with the very red 
2MASS085902.07-473209.4 (0\farcs3 offset). The observed K magnitude is 13.2 with H$-$K$_s <$ 
1.2 (the H value is an upper limit).  It is well fitted to a 9.6 keV thermal plasma with a very high column, \nh = 3.8 $\times 10^{22}\persqcm$ with a  reduced  $\chi^2$ of $0.4$.  The equivalent A$_K$ is about 2.1 which gives an absolute K magnitude of about 0.  About 50\% the flux from this star emanated during a flare.  This flare would have to be extremely hot $\sim 150$ MK to interpret this as a normal T Tauri star with a flare.

{\bf Source 108 --} This source is located just southwest of Source~149, 132\arcsec\ off axis.  Its X-ray spectrum is very similar to nearby 
Sources 147 and 149, kT= 8.9, \nh = 5.37 $\times 10^{22}\persqcm$ with a  reduced  $\chi^2$ of $0.4$.  There is no near-IR detection at this location in the 2MASS database, nor would we expect any if the source is similar to 149 and 
147 given the additional extinction present.

These stars are all interpreted as O or B stars which should, in principle, have strong, optically thick winds.  However, in the sample of 8 such stars in the ONC by Stelzer \e (2005) only 2 sources needed thermal plasma in excess of 3.33~keV to fit the data.  In both cases the hot component was a fit artifact listed as ``kT$>$ 15 keV'', in one case the hot component was less than 
10\% of the X-ray flux, in the other case it was $<$ 1\%.  We do not believe there is evidence for a strong optically thick wind giving rise to such temperatures.   The rate of flaring, 2/5 in 100 ks is also inconsistent with a collisional wind model.  This leads us to conclude that these are good candidates for objects undergoing extremely high rates of accretion.

\begin{figure}[h] 
\plotfiddle{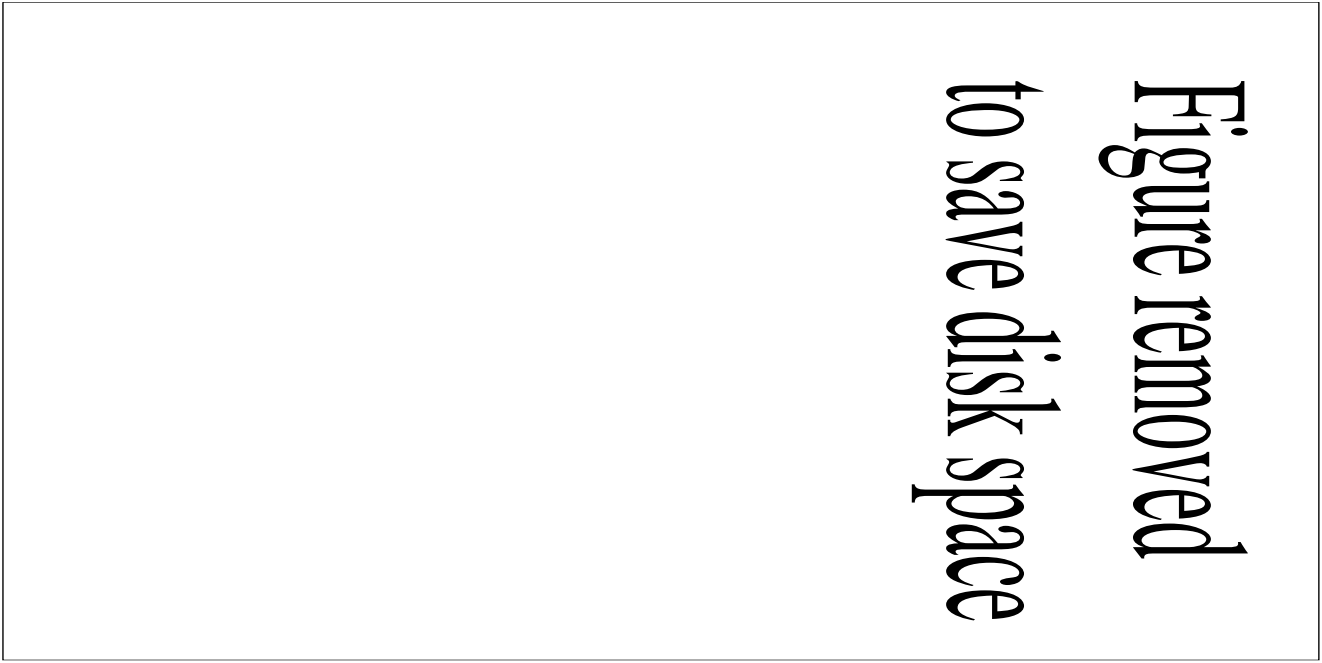}{0.0in}{-90.}{250.}{500.}{-50.}{0.}
\caption{Left: Location of the X-ray sources with hot ($>8$keV) corona or embedded with a measured column with \nh $> 10^{23}\persqcm$, overlaid on a 2MASS K$_s$ band image 8\arcmin\ on a side.  Right: Close up view of the central
2.5\arcmin.  In both figures, the hot (circles) and embedded (squares) sources are co-aligned with the cold dust indicated by the mm data (contours 
from Vigil \e in prep.).  The thick contour is 60\% of the 1.2 mm peak.  Source 
251 is both hot and embedded.}
\label{hot_embed}
\end{figure}

We note sources 251, 147, 149 and 108 lie on a nearly straight line running from northeast to southwest.  This line is parallel to the dust extension seen in the mm observation reported by Vigil \e and the axis of the diffuse plasma emission reported by Wolk \e (2002). Source 251 lies near the northern extent of the peak plasma emission (Figure~\ref{hot_embed}).  There is an additional cluster member associated with very hot plasma and over 50 counts, Source 345, which also lies close to this line. Source 345 is unusual among the hot sources because it is a fairly low extinction \nh = 7.5 $\times 10^{21}\persqcm$ and corresponds to a fairly bright (K=13.46) 2MASS source.  This is the luminosity expected from a 1.5 $M_{\odot}$ star at about 0.5 Myr.

In addition to Source 251 there in only one bright source with a measured column with \nh $> 10^{23}\persqcm$ -- Source 118.  This appears to be a normal PMS star with a hot component of about 3.7 keV and cool component of about 0.9 keV.  The spectrum is well fitted with a two-temperature Raymond--Smith plasma with a reduced $\chi^2$ of about $0.25$.  From the fit we derive a very high X-ray luminosity log L$_x= 32.9.$ The source is marginally detected by the VLT in the K-band giving it a K-band magnitude of $<$ 18.6.  Three other sources with over 50 counts were found to be absorbed by columns with \nh $>10^{23}\persqcm$ -- Sources 28, 78 and 322. These sources all lie along the same line as the stars with very hot plasma (Figure~\ref{hot_embed}). Perhaps this is a filament of gas and dust behind the currently active region of star formation.

\section{Summary} 
We observed the massive, embedded young cluster RCW 38 and the surrounding region using the $Chandra$ X-ray Observatory and in the near-IR using the VLT. 
Here we summarize our results. 

\begin{itemize}
\item   We detected 460 X-ray point sources in the field. The limiting luminosity for sources in the cluster is about log $L_x\sim 30.0$.
We detect over half of all stars down to 0.5 $M_\odot$.  
Some of the detections appear to be $<$0.1 $M_\odot$

\item Using quartile analysis we identify 360 of the X--ray sources as cluster members.  Since this approach is unusual we verify the veracity of the sources identified this way using variability as an independent tracer of stellar nature. Of the 35 stars seen to vary with 99\% confidence, 34 of them are identified as cluster members by the quartile approach.  This gives us great confidence in this method among the bright sources.  
Ten percent of the cluster members with over 30 counts are seen to flare.  This gives a flare rate of one flare per 775 ks.

\item We studied several thermal plasma models and conclude that the 
Raymond--Smith model is most appropriate for these sources.
The cluster members have a typical plasma temperature of 2.9 keV with a mean absorption of 2.8$\times 10^{22}$ cm$^{-2}$.  By comparing the infrared observed extinction with the fitted \nh\ we derive a best fit of \nh\ = A$_V$ $\times 2.0 
\times 10^{21}$ cm$^{-2}$.

\item The center of the cluster is located at 08h~59d~04.64s,
-47d~30m~44.00s (J2000) and has a peak central density of $\approx$ 400 X-ray sources parsec$^{-2}$ in the central 0.1 parsec$^{-2}$ assuming a 1.7 kpc distance of RCW~38. 
The density of the number of sources per parsec$^2$ is $\sim exp(-5.0 d)$ where $d$ is the distance in parsecs from the cluster center. The cluster is highly centrally condensed with a half-width at half maximum density of 0.2~pc. 

\item Since the exposure only reached log $L_x \sim 30$ we only detected between 15 and 25\% of the members based on comparison with the ONC. 
So the true peak density is about 1600 stars parsec$^{-2}$ in the central 0.1 parsec.  A similar extrapolation of X-ray detected members to total membership puts the total membership between 1400 and 2400.

\item We find it highly unlikely that the diffuse emission reported by
Wolk \e (2002) is the result of the 1000--2000 stars not detected as individual X-rays sources.  This is because the spectrum of the diffuse emission is distinctly non-stellar, the extent of the diffuse emission is greater than that of the cluster and there are more photons present in the diffuse emission than would be expected from the non-detected point sources.  The point sources may be responsible for the diffuse emission in the central 0.25\arcmin.

\item The KLF of the X--ray sources appears quantitatively similar to that of the non--X--ray selected sample indicating that the X-ray sample is not particularly biased except at the lowest masses. Uncertainties in the various mass tracks make stating a value for the completeness limit, in terms of stellar mass, meaningless.

\item About 150 of the X-ray cluster members are matched to near infrared sources in the 2MASS catalog or in our VLT observations of the center of the RCW~38 cluster. Less than one-third of the X-ray selected sample of cluster members with near infrared colors appears to have optically thick disks. 
In addition, there are 353 non-X-ray emitting infrared sources in the central
2.5\arcmin\ $\times$ 2.5\arcmin\ detected by the VLT.  The bulk of these (at least 90\%) are cluster members. Based on the near IR colors, at least 50\% of these non--X--ray detected infrared sources possess disks.  This indicates a bias of the X-ray sources against having disks which are optically thick at K-band.

\item We identify 31 OB star candidates in the field assuming an age of
0.5 Myr.  This number is consistent with the number of OB stars expected for a cluster with about 2000 members. If we assume a cluster age of 1~Myr, we would have identified over 60 OB stars.  This is inconsistent with the size of the cluster and is taken as evidence for the younger age. A continuum of stellar ages is probably present as there are still protostellar candidates in the cluster, in particular associated with the IRS~1 ridge.
\end{itemize}

\acknowledgments
This publication makes use of data products from the Two Micron All
Sky Survey, which is a joint project of the University of Massachusetts and the 
Infrared Processing and Analysis Center, funded by the National Aeronautics and 
Space Administration and the National Science Foundation. This work was supported by CXC guest investigator grant GO2-3103X.  We also thank the referee Leisa Townsley for many helpful comments.


\clearpage

\clearpage




\begin{deluxetable}{lcccrrcc}
\tablecaption{X-ray sources detected in the RCW 38 Field \label{Sources}}
\tablewidth{0pt}
\tablehead{
\colhead{SRC \#} &\colhead{CXOJ} &\colhead{RA}  &\colhead{DEC}   &  \colhead{Off-axis}   &
\colhead{Net counts} &  \colhead{Baysian} &\colhead{Baysian}\\
\colhead{} &\colhead{} &\colhead{(J2000.0)}  &\colhead{(J2000.0)}   & \colhead{angle}   &
\colhead{} &  \colhead{Blocks}&\colhead{Blocks}\\
\colhead{} &\colhead{} &\colhead{}  &\colhead{}   & \colhead{(\arcsec)}   &
\colhead{} &  \colhead{95\% sig.}&\colhead{99\% sig.}}
\startdata
  1&085821.3-473107&8:58:21.26&-47:31:07.2&604.6&54.1 & 1 & 1  \\     
  2&085823.8-473001&8:58:23.77&-47:30:01.8&577.1&38.5 & 1 & 1  \\     
  3&085827.6-473009&8:58:27.59&-47:30:09.2&538.3&68 & 1 & 1  \\       
  4&085829.9-472824&8:58:29.89&-47:28:24.1&526.6&96.6 & 1 & 1  \\     
  5&085830.9-473221&8:58:30.87&-47:32:21.2&520.6&28 & 1 & 1  \\       
  6&085831.5-472808&8:58:31.51&-47:28:08.1&514.3&60.3 & 1 & 1  \\     
  7&085833.1-473152&8:58:33.15&-47:31:52.6&491.8&24.1 & 1 & 1  \\     
  8&085834.6-472816&8:58:34.61&-47:28:16.5&481.8&119.3 & 2 & 1  \\
\enddata                                         
\end{deluxetable}

\begin{deluxetable}{lcccccccc}
\tablecaption{Quartile Values of Probable Cluster Members \label{quantin}}
\tablewidth{0pt}
\tablehead{
\colhead{SRC\#} &\colhead{Q25} &\colhead{Q25$_{err}$} &\colhead{M}&
\colhead{M$_{err}$}  & \colhead{Q75}   &\colhead{Q75$_{err}$} &
 \colhead{log(m/(1-m))} &\colhead{3*Q25/Q75} }
\startdata
 4&0.15&0.01&0.23&0.02&0.33&0.04&-0.53&1.32	 \\
 8&0.18&0.01&0.29&0.03&0.45&0.03&-0.39&1.17	 \\
 11&0.27&0.05&0.41&0.03&0.57&0.05&-0.16&1.41	 \\
 16&0.19&0.02&0.31&0.03&0.47&0.09&-0.35&1.24	 \\
 19&0.17&0.01&0.26&0.06&0.57&0.06&-0.46&0.90	 \\
 21&0.16&0.01&0.22&0.01&0.39&0.02&-0.55&1.19	 \\
 22&0.31&0.03&0.45&0.06&0.71&0.13&-0.08&1.31	 \\
 23&0.14&0.01&0.22&0.01&0.34&0.03&-0.56&1.20	 \\
\enddata
\end{deluxetable}

\begin{deluxetable}{lcccccccc}
\tablecaption{Quartile Values of Probable Non-Members \label{quantout}}
\tablewidth{0pt}
\tablehead{
\colhead{SRC\#} &\colhead{Q25} &\colhead{Q25$_{err}$} &\colhead{M}&
\colhead{M$_{err}$}  & \colhead{Q75}   &\colhead{Q75$_{err}$} &
 \colhead{log(m/(1-m))} &\colhead{3*Q25/Q75} }
\startdata
1&0.11&0.01&0.16&0.02&0.44&0.14&-0.74&0.74      \\
2&0.20&0.02&0.37&0.08&0.95&0.13&-0.23&0.62      \\
3&0.19&0.01&0.29&0.04&0.60&0.11&-0.39&0.95      \\
5&0.19&0.03&0.41&0.09&0.68&0.19&-0.15&0.85      \\
6&0.11&0.01&0.16&0.03&0.35&0.14&-0.71&0.92      \\
7&0.31&0.06&0.68&0.08&1.16&0.13&0.32&0.80       \\
9&0.20&0.03&0.34&0.14&1.10&0.21&-0.28&0.53      \\
10&0.20&0.03&0.37&0.05&0.71&0.28&-0.24&0.84     \\
\enddata
\end{deluxetable}

\begin{deluxetable}{lrrrrrrr}
\tablecaption{Summary Model Fit Data \label{test-res}}
\tablewidth{0pt}
\tablehead{
\colhead{Model} &\colhead{RS}  &\colhead{2temp RS}   & \colhead{APEC}   &
\colhead{MeKaL} &  \colhead{TB}  & \colhead{BB} & \colhead{Power Law}}
\startdata
mean     & 0.41 &0.44 & 0.41&  0.45&  0.44& 1.44& 1.23\\
MAD & 0.017&0.019 & 0.016& 0.020&  0.019& 0.090 & 0.069\\
Num. Rej. &  14  & 9   &  19 &  32  & 16   & 16  &   11\\
\enddata
\end{deluxetable}

\begin{deluxetable}{lcccccccccc}
\tabletypesize{\footnotesize}
\tablecaption{Sources with Two--Temperature Spectra \label{2t}}
\tablewidth{0pt}
\tablehead{ \colhead{SRC\#}  &\colhead{$\chi^2$} & 
\colhead{\nh} &  \colhead{\nh$_{err}$} & \colhead{kT$_1$}  & 
 \colhead{kT$_{1~err}$} & \colhead{kT$_2$}  &  \colhead{kT$_{2~err}$}&
\colhead{Unabs. Flux} & \colhead{EMR$^a$} &  \colhead{L$_{x}$} \\
\colhead{} & \colhead{per d.o.f.} & 
\colhead{10$^{22}~cm^{-2}$} &  \colhead{} & \colhead{keV}  &
\colhead{}  & \colhead{keV} &  \colhead{} &\colhead{log ergs cm$^{-2}\ps$} & 
\colhead{kT$_1$/kT$_2$} &  \colhead{log ergs$\ps$}}
\startdata
8  &0.28&2.30&0.26&1.37&0.50&6.47&--&-13.23&2.79&31.31      \\
23 &0.47&1.24&0.12&0.60&0.34&3.04&1.45&-13.31&1.61&31.23    \\
39 &0.10&0.67&0.02&0.08&0.00&0.35&0.06&-11.31&32.41&b	    \\
97 &0.29&1.78&0.21&0.49&--&4.02&--&-13.29&0.12&31.25	    \\
104&0.39&2.88&0.14&0.31&0.08&5.67&3.85&-11.82&54.04&32.72   \\
118&0.24&21.36&1.70&0.90&0.18&3.67&3.11&-11.64&12.65&32.90  \\
129&0.42&3.50&0.24&1.44&0.20&$>$15&--&-12.80&8.70&31.74	    \\
133&0.12&1.25&0.19&0.47&--&3.74&0.00&-13.54&0.07&31.00	    \\
\enddata
\end{deluxetable}

\begin{deluxetable}{rrcccccc}
\tablecaption{Cluster Members Fit  with 1 Temperature Spectra \label{1t}}
\tablewidth{0pt}
\tablehead{ 
\colhead{SRC\#}  &\colhead{$\chi^2$} & \colhead{\nh} &
\colhead{\nh$_{err}$}  &\colhead{kT} &  \colhead{kT$_{err}$} &
\colhead{Unabs. Flux} & \colhead{L$_{x}$} \\ 
\colhead{}  & \colhead{per d.o.f.} & 
\colhead{10$^{22}~cm^{-2}$} &\colhead{}& \colhead{keV} &\colhead{}&
\colhead{log ergs cm$^{-2}\ps$} & \colhead{log ergs$\ps$}}
\startdata
3  &0.23&3.19&1.06&1.04&0.46&-13.08&31.45     \\
4  &0.36&1.22&--&1.72&--&-13.42&31.11	  \\
11 &0.41&3.54&0.63&3.72&--&-13.43&31.11	  \\
14 &0.17&4.80&1.79&3.51&1.16&-13.60&30.94	  \\
16 &0.19&2.22&1.37&2.08&1.59&-13.69&30.85	  \\
19 &1.01&2.48&1.14&0.73&0.40&-13.28&31.26	  \\
21 &0.48&0.90&0.16&3.87&1.09&-13.17&31.36	  \\
22 &0.32&4.46&1.04&3.67&--&-13.54&30.99	  \\
\enddata					  
\end{deluxetable}

\begin{deluxetable}{cccccccc}
\tablecaption{Faint Cluster Members Fit  with 1 Temperature Spectra 
\label{1t-faint}}
\tablewidth{0pt}
\tablehead{ 
\colhead{SRC\#}  &\colhead{$\chi^2$} & \colhead{\nh} &
\colhead{\nh$_{err}$}  &\colhead{kT} &  \colhead{kT$_{err}$} &
\colhead{Unabs. Flux} & \colhead{L$_{x}$} \\ 
\colhead{}  & \colhead{per d.o.f.} & 
\colhead{10$^{22}~cm^{-2}$} &\colhead{}& \colhead{keV} &\colhead{}&
\colhead{log ergs cm$^{-2}\ps$} & \colhead{log ergs$\ps$}}
\startdata
26 &0.26&3.39&0.94&2.06&0.63&-13.84&30.00        \\
27 &0.26&8.80&1.12&0.46&0.05&-11.64&29.88	     \\
38 &0.27&1.39&0.57&$>$15&--&-13.98&30.39     \\
41 &0.22&16.87&3.19&0.87&0.12&-12.22&29.98	     \\
43 &0.31&2.03&0.55&3.01&1.05&-13.85&30.20	     \\
45 &0.24&2.12&0.73&4.70&1.99&-13.96&30.21	     \\
47 &0.27&7.15&1.74&2.09&0.49&-13.41&30.21	     \\
48 &0.30&0.85&0.29&1.88&0.59&-14.07&29.98	     \\
\enddata
\end{deluxetable}

\begin{deluxetable}{lccccccc}
\tablecaption{X-ray Spectral fits to Bright Non-member X-ray Sources. 
\label{non}}
\tablewidth{0pt}
\tablehead{ 
\colhead{SRC\#} &\colhead{$\chi^2$} & \colhead{\nh} &
\colhead{\nh$_{err}$}  &\colhead{kT} &  \colhead{kT$_{err}$} &
\colhead{Unabs. Flux}  \\ 
\colhead{} & \colhead{per d.o.f.} & 
\colhead{10$^{22}~cm^{-2}$} &\colhead{}& \colhead{keV} &\colhead{}&
\colhead{log ergs cm$^{-2}\ps$} }
\startdata
1  &0.54&1.24&0.42&0.34&0.08&-13.00        \\
2  &0.14&2.43&--&1.82&--&-13.70    \\
6  &0.20&1.00&0.39&0.72&0.29&-13.57	       \\
13 &0.34&0.46&0.34&4.60&0.15&-14.22       \\
32 &0.62&0.26&0.37&4.28&0.13&-14.41       \\
34 &0.12&0.00&0.17&0.30&--&-14.46     \\
54 &0.06&0.01&--&1.04&--&-14.32   \\
61 &0.21&4.82&0.66&3.31&--&-13.72     \\
64 &0.18&0.53&0.48&0.76&0.44&-14.19       \\
69 &0.53&0.60&0.15&0.37&0.07&-13.11      \\
80 &0.35&0.51&--&0.82&--&-14.17   \\
113&0.50&0.56&0.43&2.46&1.17&-14.09      \\
175&0.16&0.64&0.37&0.90&0.27&-14.10      \\
214&0.37&0.86&0.11&4.38&0.76&-13.17     \\
245&0.09&1.02&--&1.76&--&-14.01  \\
277&0.16&0.16&--&0.91&--&-14.50  \\
286&0.30&0.40&--&0.60&--&-14.21  \\
399&0.34&1.25&--&1.14&--&-13.39 \\
434&0.08&0.75&0.03&1.89&--&-14.22    \\
446&0.37&0.49&0.25&1.00&0.19&-13.63      \\
447&0.40&1.18&0.12&0.29&0.04&-11.95     \\
452&0.24&0.75&0.21&0.14&0.03&-12.30      \\
454&0.22&1.84&0.62&0.64&0.27&-13.37      \\
\enddata
\end{deluxetable}

\begin{deluxetable}{lcccccc}
\tablecaption{X-ray Spectral fits to Faint Non-member Sources. \label{nonpoor}}
\tablewidth{0pt}
\tablehead{
\colhead{SRC\#}  &\colhead{$\chi^2$} & \colhead{\nh} &
\colhead{\nh$_{err}$}  &\colhead{kT} &  \colhead{kT$_{err}$} &
\colhead{Unabs. Flux}  \\ 
\colhead{}  & \colhead{per d.o.f.} & 
\colhead{10$^{22}~cm^{-2}$} &\colhead{}& \colhead{keV} &\colhead{}&
\colhead{log ergs cm$^{-2}\ps$} }
\startdata
5  &0.37&0.27&--&$>$15&--&-14.15      \\
7  &0.40&3.26&1.27&$>$15&--&-13.82      \\
9  &0.33&6.69&0.85&0.39&0.03&-11.72	       \\
10 &0.35&3.32&0.64&1.27&0.27&-13.44	       \\
12 &0.33&2.97&0.97&2.80&1.06&-13.85	       \\
14 &0.35&3.64&1.12&3.71&1.28&-13.67	       \\
15 &0.25&0.63&0.15&0.34&0.06&-13.87	       \\
17 &0.25&1.28&0.14&0.17&0.02&-12.02	       \\
\enddata			  
\end{deluxetable}


\begin{deluxetable}{rrcccccccccc}
\tabletypesize{\tiny}
\tablecaption{IR Photometry of X-ray Sources (members)
\label{tbl-irmembers}} 
\tablewidth{0pt}
\tablehead{ \colhead{SRC\#} & \colhead{IR ID} & \colhead{offset} & 
\colhead{J} &  \colhead{Jerr}  & \colhead{H} & \colhead{Herr} & 
\colhead{K} &  \colhead{Kerr}  & \colhead{A$_V$} & \colhead{Model} &
\colhead{flags}}
\startdata
  4&2MASS085829.85-472823.0&1.11&14.058&0.028&12.895&0.026&12.429&0.019&5.16&GK STAR&AAA      \\ 
  8&2MASS085834.64-472816.4&0.30&14.968&0.032&13.312&0.02&12.314&0.019&10.32&M STAR&AAA   \\ 
 14&2MASS085842.04-472628.3&1.29&15.459&0.047&14.406&0.054&13.98&0.06&4.4&Bright&AAA	      \\ 
 16&2MASS085842.58-473156.7&0.91&14.599&0.033&13.602&0.046&13.12&0.036&2.74&Bright&AAA	      \\ 
 19&2MASS085843.51-472934.3&0.40&16.556&0.161&14.881&0.062&14.049&0.068&11.14&Bright&CAA     \\ 
 21&2MASS085843.88-472856.9&0.22&12.572&0.023&11.529&0.027&10.76&0.023&5.35&M STAR&AAA    \\ 
 22&2MASS085844.19-473441.4&1.44&17.086&---&15.336&0.113&13.536&0.067&---&---&UBA	      \\ 
 23&2MASS085844.40-473253.3&0.26&14.64&0.042&13.299&0.03&12.658&0.035&6.22&Disk&AAA	      \\
\enddata
\end{deluxetable}

\begin{deluxetable}{rrrrrrrrrrrr}
\tabletypesize{\footnotesize}
\tablecaption{IR Photometry of X-ray Sources (non-members) \label{tbl-irnon}}
\tablewidth{0pt}
\tablehead{ \colhead{SRC\#} & \colhead{IR ID} & \colhead{offset} & 
\colhead{J} &  \colhead{Jerr}  & \colhead{H} & \colhead{Herr} & 
\colhead{K} &  \colhead{Kerr}  & \colhead{A$_V$} & \colhead{Model} &
\colhead{flags}}
\startdata
  1&2MASS085821.23-473107.9&0.73&15.154&0.045&14.123&0.032&13.879&0.049&5.12&GK 
STAR&AAA     \\
  2&2MASS085823.82-473001.6&0.52&16.171&0.12&14.048&0.065&12.772&0.048&11.94&GK 
STAR&BAA     \\
  3&2MASS085827.50-473008.8&0.99&16.532&0.121&14.926&0.047&14.095&0.053&11.64&Bright&BAA     \\ 
  6&2MASS085831.46-472807.9&0.52&14.047&0.031&13.389&0.032&13.253&0.038&0.82&GK 
STAR&AAA     \\
 10&2MASS085836.16-473135.8&0.60&17.356&---&15.527&0.116&14.504&0.085&---&---
&UBA	      \\
 12&2MASS085839.69-473127.1&1.03&17.169&---&15.011&0.057&13.754&0.042&---&---
&UAA	      \\
 13&2MASS085841.58-472835.0&0.34&15.577&0.071&14.273&0.043&13.743&0.051&6.6&GK 
STAR&AAA      \\
 15&2MASS085842.42-473041.6&1.34&12.376&0.023&11.98&0.02&11.922&0.026&0&GK 
STAR&AAA	      \\
\enddata
\end{deluxetable}


\begin{deluxetable}{rrrrrrrrrrrrr}
\tabletypesize{\small}
\tablecaption{IR Photometry in the VLT field (Non--X--ray Sources) \label{tbl-irnox}}
\tablewidth{0pt}
\tablehead{ \colhead{IR ID} & \colhead{RA} & \colhead{DEC} & 
\colhead{J} &  \colhead{Jerr}  & \colhead{H} & \colhead{Herr} & 
\colhead{K} &  \colhead{Kerr}& \colhead{A$_V$}\\
\colhead{} & \colhead{J2000.} & \colhead{J2000.} & 
\colhead{} &  \colhead{}  & \colhead{} & \colhead{} & 
\colhead{} &  \colhead{} &\colhead{min.} }
\startdata
VLT085911.02-472949.8&  08:59:11.022& -47:29:49.81&   19.537&    0.148&   17.724&    0.048&   16.190 &   0.038 &    5.58\\ 
VLT085910.86-473101.6&  08:59:10.867& -47:31:01.67&   16.763&    0.033&   16.097&    0.020&   15.586 &   0.023 &    0.00\\ 
VLT085910.79-473028.9&  08:59:10.795& -47:30:28.94&   18.031&    0.068&   16.531&    0.025&   14.550 &   0.015 &    0.00\\ 
VLT085910.72-473110.4&  08:59:10.720& -47:31:10.40&   19.667&    0.149&   17.745&    0.046&   16.176 &   0.034 &    6.81\\ 
VLT085910.59-472945.7&  08:59:10.596& -47:29:45.70&   18.288&    0.068&   16.753&    0.027&   15.356 &   0.021 &    2.84\\ 
VLT085910.40-472933.7&  08:59:10.405& -47:29:33.72&   18.673&    0.087&   17.420&    0.038&   16.060 &   0.032 &    0.00\\ 
VLT085910.38-472942.6&  08:59:10.385& -47:29:42.61&   17.808&    0.054&   15.600&    0.016&   13.753 &   0.010 &    8.54\\ 
VLT085910.08-473133.8&  08:59:10.088& -47:31:33.87&   17.889&    0.056&   16.588&    0.025&   15.528 &   0.023 &    2.30\\
\enddata 
\end{deluxetable}

\begin{deluxetable}{rrrrrrrrrrrr}
\tabletypesize{\small}
\tablecaption{Candidate OB stars in RCW~38 \label{tbl-ob}}
\tablewidth{0pt}
\tablehead{ \colhead{SRC \#} & \colhead{RA} & \colhead{DEC} & 
\colhead{off--axis} &  \colhead{A$_V$} & 
\colhead{A$_V$} &  \colhead{Abs. K} &  \colhead{L$_{bol}$}& 
\colhead{L$_X$/L$_{bol}$}\\
\colhead{} & \colhead{(J2000.0)} & \colhead{(J2000.0)} & 
\colhead{(angle)} &  \colhead{(From IR)} & \colhead{(From \nh)} &  \colhead{} 
&
\colhead{(log $L_\odot$)} & \colhead{(log)}  \\
\colhead{} & \colhead{} & \colhead{} & 
\colhead{(arcsec)} &  \colhead{(mag)} & \colhead{(mag)} &  \colhead{} 
& \colhead{} & \colhead{} }
\startdata
21 &8:58:43.87&-47:28:57.1&236.1     &1.1&4.5   &-0.51   &3.09&-5.32  \\
53 &8:58:50.37&-47:33:19.4&108.9     &8.9&8.9   &-1.81   &3.62&-5.48  \\
107&8:58:59.20&-47:26:57.9&232.7     &4.4&12.2  &-1.31  &3.42&-6.02  \\
112&8:58:59.79&-47:32:23.5&111.0     &23.9&26.9 &-4.35 &4.64&-7.71   \\
125&8:59:00.33&-47:31:14.3&53.1      &31.5&31.2 &-2.17 &3.76&-6.32   \\
127&8:59:00.39&-47:25:58.3&288.9     &13&17.0   &-1.97   &3.68&-6.34   \\
140&8:59:01.56&-47:31:10.2&40.7      &8.7&14.9  &-0.65  &3.15&-5.50    \\
143&8:59:01.64&-47:30:37.5&31.0      &2.7&5.8   &-1.02   &3.30&-5.96   \\
\enddata 
\end{deluxetable} 
\end{document}